\documentclass[11pt]{article}
\pdfoutput=1 
\usepackage{jheppub}
\usepackage{bbm,bm,mathtools,slashed}
\usepackage{multirow}
\usepackage[T1]{fontenc}
\usepackage[nottoc]{tocbibind}
\usepackage[toc,page]{appendix}
\usepackage{youngtab}
\usepackage{physics}

\graphicspath{{./figs/}}

\newcommand{\ri}{\mathrm{i}}
\newcommand{\ER}{{R}}
\newcommand{\EL}{{L}}

\newcommand{\be}{\begin{equation}}
\newcommand{\ee}{\end{equation}}
\newcommand{\bea}{\begin{eqnarray}}
\newcommand{\eea}{\end{eqnarray}}

\newlength{\fighskip} \fighskip=2pt
\newlength{\figvskip} \figvskip=3pt

\preprint{KEK-TH-2525, J-PARC-TH-0286,RIKEN-iTHEMS-Report-23}
\title{String-net formulation of Hamiltonian lattice Yang-Mills theories and quantum many-body scars in a nonabelian gauge theory}

\author[a,b]{Tomoya Hayata,}
\author[b,c,d,e,f]{Yoshimasa Hidaka}

\affiliation[a]{Departments of Physics, Keio University, 4-1-1 Hiyoshi, Kanagawa 223-8521, Japan}
\affiliation[b]{RIKEN iTHEMS, RIKEN, 2-1, Hirosawa, Wako, Saitama 351-0198, Japan}
\affiliation[c]{Theory Center, Institute of Particle and Nuclear Studies, High Energy Accelerator Research Organization (KEK), 1-1 Oho, Tsukuba 305-0801, Japan}
\affiliation[d]{Graduate Institute for Advanced Studies, SOKENDAI, 1-1 Oho, Tsukuba 305-0801, Japan}
\affiliation[e]{nternational Center for Quantum-field Measurement Systems for Studies of the Universe and Particles (QUP), KEK, 1-1 Oho, Tsukuba 305-0801, Japan}
\affiliation[f]{Department of Physics, Faculty of Science, University of Tokyo, 7-3-1 Hongo Bunkyo-ku Tokyo 113-0033, Japan}

\emailAdd{hayata@keio.jp}
\emailAdd{hidaka@post.kek.jp}

\abstract{
We study the Hamiltonian lattice Yang-Mills theory based on spin networks that provide a useful basis to represent the physical states satisfying the Gauss law constraints.
We focus on $\mathrm{SU}(2)$ Yang-Mills theory in $(2+1)$ dimensions.
Following the string-net model, we introduce a regularization of the Kogut-Susskind Hamiltonian of lattice Yang-Mills theory based on the $q$ deformation, which respects the (discretized) $\mathrm{SU}(2)$ gauge symmetry as quantum group, i.e., $\mathrm{SU}(2)_k$, and enables implementation of the lattice Yang-Mills theory both in classical and quantum algorithms by referring to those of the string-net model.
Using the regularized Hamiltonian, we study quantum scars in a nonabelian gauge theory.
Quantum scars are nonthermal energy eigenstates arising in the constrained quantum many-body systems.
We find that quantum scars from zero modes, which have been found in abelian gauge theories arise even in a nonabelian gauge theory.
We also show the spectrum of a single-plaquette model for SU(2)$_k$ and SU(3)$_k$ with naive cutoff and that based on the $q$-deformation to discuss cutoff dependence of the formulation.

}

\begin{document}
\maketitle

\section{Introduction}

Lattice gauge theories (LGTs) are one of the most established and powerful formulations of quantum field theories, which have been developed for nonperturbative calculations of quantum chromodynamics (QCD) to understand the physics of strong interactions from first principles~\cite{Wilson:1974sk}.
Recently, LGTs have caught the interest of researchers in fields outside of high-energy physics~\cite{cirac_goals_2012}.
It was found that LGTs are good playgrounds to test the performance of two new complementary approaches. One is a classical simulation based on tensor network methods~\cite{Orus:2013kga,Banuls:2019rao,Banuls:2019bmf}, and the other is a quantum computer or simulator~\cite{Preskill:2018fag,Zohar:2021nyc}. Even experimental realizations of LGTs have been challenged, e.g., in cold atomic systems~\cite{Dalmonte:2016alw}.
These approaches are considered as a promising method for simulating LGTs without suffering from the notorious sign problem~\cite{Troyer:2004ge}, which is necessary to understand QCD phase diagram at finite density or to solve real-time problems of QCD such as the dynamical formation of quark-gluon-plasma in heavy-ion collision experiments.
Not only such computational challenges of LGTs as targets for a potential quantum advantage, but also LGTs themselves have been intriguing from quantum statistical mechanics and quantum information perspectives.
For example, quantum statistical problems such as ergodicity breaking (violation of eigenstate thermalization hypothesis) in kinematically constrained systems have been actively studied in recent years. 
Nonthermal energy eigenstates arising in the constrained quantum many-body systems are referred to as quantum many-body scars~\cite{bernien_probing_2017,turner_weak_2018,2020PhRvX..10a1047S,Khemani:2019vor,Desaules:2022ibp,Desaules:2022kse,Su:2022glk} (see ref.~\cite{Serbyn:2020wys} for a review).
Indeed, Hamiltonian LGTs are prototypes of such constrained quantum many-body systems since the gauge invariance, that is, the Gauss law is imposed as the constraint conditions to the Hilbert space~\cite{Brenes:2017wzd,2019PhRvR...1c3144O,Karpov:2020nhy,Banerjee:2020tgz,Biswas:2022env}.

Hamiltonian LGTs, which were pioneered by Kogut and Susskind~\cite{Kogut:1974ag}, are used in quantum computations instead of the conventional path integral formulation~\cite{Klco:2019evd,Atas:2021ext,ARahman:2021ktn,Hayata:2021kcp,Ciavarella:2021nmj,Davoudi:2022xmb,Yao:2023pht}.
To implement the Kogut-Susskind Hamiltonian on tensor networks or quantum computers, which can handle only finite-dimensional Hilbert space, we need to introduce a cutoff to gauge fields with keeping gauge invariance manifestly since gauge fields have infinite-dimensional Hilbert space even after the continuum theory is regularized on a finite lattice.
Although several formulations have been developed so far, which can be classified by the computational basis and cutoff scheme (see e.g., ref.~\cite{Davoudi:2020yln}), the search for an efficient formulation that has better cutoff dependence and meets quantum hardware requirements flexibly is still actively studied.

In this paper, we formulate a regularized Hamiltonian lattice Yang-Mills theory based on spin networks~\cite{penrose1971angular,Rovelli:1995ac,Baez:1994hx,Burgio:1999tg}, in a form that can be computed numerically.
Using it, we study quantum scars in a nonabelian gauge theory.
The remainder of this paper is organized as follows. In section~\ref{sec:KS}, we review the Kogut-Susskind Hamiltonian, and then formulate it on the basis of the spin networks. Regularization of the theory based on the $q$-deformation is discussed in section~\ref{quantum_deformation}.
We dub this regularization the string-net formulation since such a formulation of a nonabelian gauge theory based on quantum group in a numerically computable spin model was originally discussed in the string-net model. It was developed by Levin and Wen to construct a solvable model of topological order~\cite{Levin:2004mi}. Since this formulation has the same computational basis as the string-net model, the efficient implementation of Yang-Mills theory may be possible both in classical and quantum algorithms using it.
As another advantage of the formulation based on quantum group, the underlying lattice becomes topological. Namely, the theory is independent of a choice of trivalent graphs.
In section~\ref{sec:ED}, we study the full spectrum of SU(2)$_k$ Yang-Mills theory using exact diagonalization. In particular, we study quantum scars, which are energy eigenstates arising in the mid-range of the spectrum and violate the eigenstate thermalization hypothesis.
We found that quantum scars from zero modes~\cite{Banerjee:2020tgz,Biswas:2022env} arise even in a nonabelian gauge theory.
Section~\ref{sec:conclusions} is devoted to conclusions. In appendix~\ref{sec:appendix}, we show the spectrum of a single-plaquette model for SU(2)$_k$ and SU(3)$_k$ with naive cutoff and that based on the $q$-deformation to discuss cutoff dependence of the formulation.

\section{Hamiltonian formulation}
\label{sec:KS}
Sections~\ref{sec:KS_formulation} and~\ref{sec:square_lattice} review the Kogut-Susskind Hamiltonian formulation of $\mathrm{SU}(2)$ gauge theory~\cite{Kogut:1974ag} based on the spin networks.
In section~\ref{sec:KS_formulation}, we introduce the gauge-invariant physical space spanned by states called the spin network~\cite{penrose1971angular,Rovelli:1995ac,Baez:1994hx,Burgio:1999tg}, and show the action of the Hamiltonian on the spin networks~\cite{Robson:1981ws}.
We treat trivalent graphs such as honeycomb lattice for a technical reason. Extensions to multivalent graphs will be discussed in section~\ref{sec:square_lattice}.
The dimension of the Hilbert space is infinite, even on a finite graph.
In order to perform numerical calculations, it is necessary to approximate the Hilbert space in finite dimensions with manifestly keeping the gauge symmetry.
To this end, we employ quantum group deformations as the approximation~\cite{Bimonte:1996fq,Burgio:1999tg}, which will be explained in section~\ref{quantum_deformation}.
We express the Kogut-Susskind Hamiltonian of a regularized gauge theory explicitly using the spin networks, which is one of our main results.
\begin{figure}[h]  \label{fig:HoneycombLattice}
  \centering 
  \includegraphics[scale=0.4]{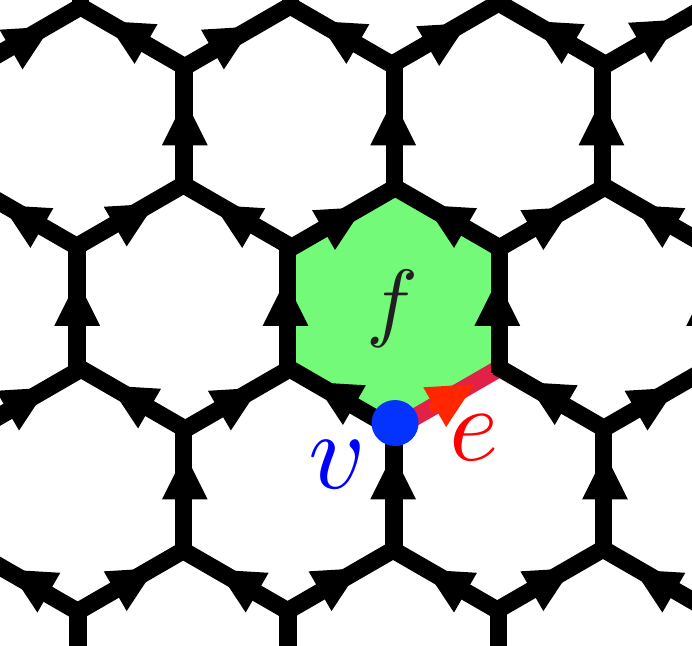}
  \caption{Honeycomb lattice. $v$, $e$, and $f$ represent a vertex, an edge, and a closed loop associated with a surface, respectively.}
\end{figure}
\subsection{Kogut-Susskind Hamiltonian formulation}\label{sec:KS_formulation}
We consider the Kogut-Susskind Hamiltonian formulation of $\mathrm{SU}(2)$ gauge theory~\cite{Kogut:1974ag} on a trivalent directed graph $\Gamma(\mathcal{V}, \mathcal{E})$,
where $\mathcal{V}$ and $\mathcal{E}$ are sets of vertices and edges.
In our numerical calculations in section~\ref{sec:ED}, we will consider a honeycomb lattice as shown in figure~\ref{fig:HoneycombLattice}.
Note that we introduce directed edges for convenience in defining operators on the graph, but the obtained results do not depend on their orientations.

The Hamiltonian of $\mathrm{SU}(2)$ lattice gauge theory is given by
\begin{equation}
  H = \frac{1}{2}\sum_{e\in {\mathcal{E}}} E_i^2(e) - K \sum_{f\in \mathcal{F}} \tr U (f).
  \label{eq:Hamiltonian}
\end{equation}
Here, $E_i^2(e)$ is the square of the electric fields defined on an edge $e\in \mathcal{E}$.
$K$ and $\tr U (f)$ are the coupling constant and the Wilson loop on a closed loop $f\in \mathcal{F}$, where $\mathcal{F}$ is the set of minimal closed loops (plaquettes). 
The closed loop can be expressed by a sequence of edges, $e_1e_2e_3\cdots e_n$,
where the head of an edge coincides with the tail of the next edge, i.e.,
$\mathrm{dst}(e_{k})=\mathrm{src}(e_{k+1})$ for $1\leq k <n$, and $\mathrm{dst}(e_{n})=\mathrm{src}(e_{1})$. Here, $\mathrm{dst}(e)$ and $\mathrm{src}(e)$ are functions that take the head and the tail from an edge, respectively.
For example, consider the following closed path $f$:
\begin{equation}
  \parbox{3.cm}{\includegraphics[scale=0.4]{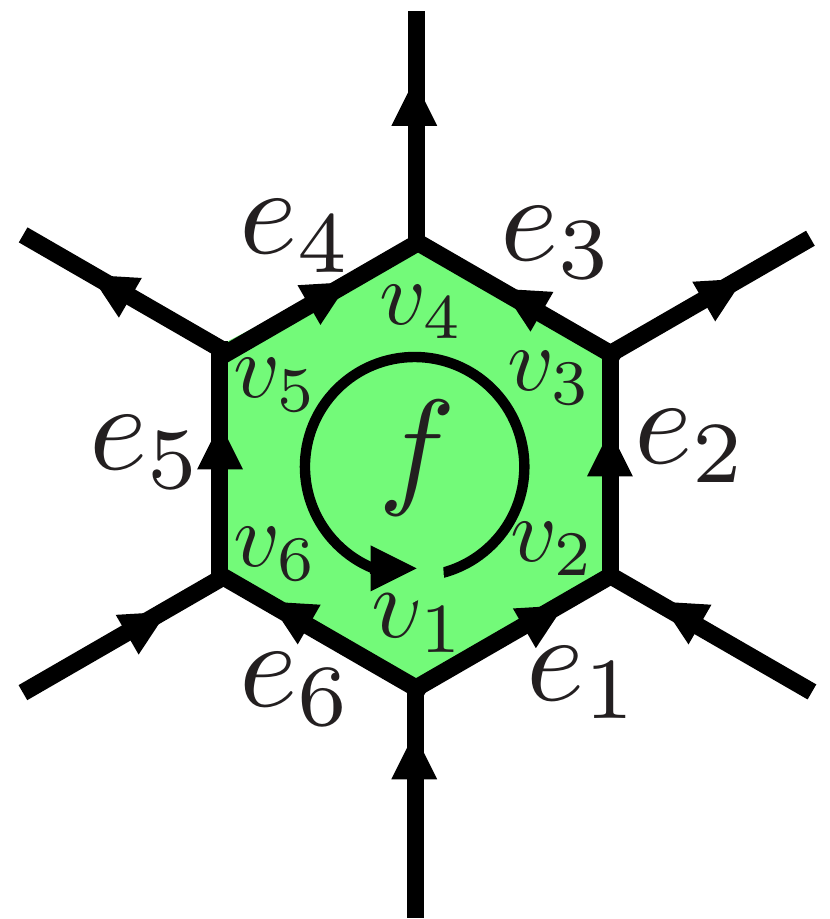}}.
  \label{eq:closed_path}
\end{equation}
The path can be expressed as $e_1e_2e_3\bar{e}_4 \bar{e}_5 \bar{e}_6$.
Here, $\bar{e}$ represents the edge with the head and tail vertices reversed.
Functions $\mathrm{dst}(e)$ and $\mathrm{src}(e)$ for these edges are
$\mathrm{src}(e_k)=v_{k}$, $\mathrm{dst}(e_k)=v_{k+1}$, 
$\mathrm{dst}(\bar{e}_k)=v_{k+1}$, and $\mathrm{src}(\bar{e}_k)=v_{k}$, respectively.

$U (f)$ is defined by the path-ordered product of link variables $U(e)$ on an edge, which is an operator-valued $2\times 2$ unitary matrix:
\begin{equation}
U (f) = \tr U^\dag(e_6)U^\dag(e_5)U^\dag(e_4)U(e_3)U(e_2)U(e_1).
\end{equation}
Here, we employ $U(\bar{e}_i)=U^\dag(e_i)$.
Since $\tr U (f)$ includes the trace, it does not depend on the choice of the base point of the path.

In the Hamiltonian formulation on a lattice, there are two types of electric fields, $\ER_i(e)$ and $\EL_i(e)$, that generate the gauge transformation of $U$ from the
head (left) and tail (right) sides of the edges, respectively. Their commutation relations are given by
\begin{align}
  [R_{i}(e),U(e')] &=U(e') T_i\delta_{e,e'}, \label{eq:[E_s,U]field}\\
  [L_{i}(e),U(e')] &=T_iU(e)\delta_{e,e'} , \label{eq:[E_t,U]field}
  \\
[R_{i}(e),R_{j}(e')] &=\ri f^{k}_{~ij}R_{k}(e)\delta_{e,e'},    \label{eq:[E_s,E_s]field} \\
[L_{i}(e),L_{j}(e')] &=-\ri f^k_{~ij}L_{k}(e)\delta_{e,e'},   \label{eq:[E_t,E_t]field}
\end{align}
and others vanish.
Here, $f^{k}_{~ij}=\epsilon_{ijk}$ is the structure constant of $\mathrm{SU}(2)$ with the Levi-Civita tensor $\epsilon_{ijk}$. $T_i$ is the generator of the fundamental representation satisfying $[T_i,T_j]=\ri f^k_{~ij}T_k$.
Note that $\ER_i(n)$ and $\EL_i(n)$ are not independent but related by a parallel transport:
\begin{equation}
  \ER_i(e) =\EL_j(e)[U_{\mathrm{adj}}(e)]_i^j,
  \label{eq:E_relation}
\end{equation}
where $[U_{\mathrm{adj}}(e)]_i^j$ is the link variable with the adjoint representation defined by $U T_i U^\dag=T_j[U_{\mathrm{adj}}]_i^j$.
From eq.~\eqref{eq:E_relation}, we can see the square of $\ER_i(e)$ and $\EL_i(e)$ are identical, i.e.,
$(\ER_i(e))^2=(\EL_i(e))^2\eqqcolon E_i^2(e)$. Therefore, the Hamiltonian is independent of the choice of electric fields.

We choose a local basis (representation basis) on an edge $e$, $\ket*{j_e,m_e,n_e}$, by eigenstates of commuting operators $\{R_i^2(e), R_3(e),L_3(e)\}$~\cite{Robson:1981ws,Zohar:2014qma}:
\begin{align}
  R_i^2(e)\ket*{j_e,m_e,n_e}&=C_2(j_e)\ket*{j_e,m_e,n_e}\label{eq:Risq},\\
  R_3(e)\ket*{j_e,m_e,n_e}&=n_e\ket*{j_e,m_e,n_e},\\
  L_3(e)\ket*{j_e,m_e,n_e}&=m_e\ket*{j_e,m_e,n_e},\label{eq:EL3_ket}
\end{align}
where $C(j_e)= j_e(j_e+1)$ is the quadratic Casimir invariant of representation $j_e$.
As we will see below, this basis is a convenient basis for solving the Gaussian constraints.
The total Hilbert space is spanned by  $\prod_{e\in\mathcal{E}}\ket*{j_e,m_e,n_e}$,
which contains unphysical states.
The physical Hilbert space is a subspace of the total Hilbert space constrained by the Gauss-law constraint on each vertex. A state in the physical space $\ket{\Psi}$ satisfies
\begin{equation}
  G_i(v)\ket{\Psi}=0
  \label{eq:physical_state_condition}
\end{equation}
with generators of gauge transformations,
\begin{equation}
  G_i(v)\coloneqq  \sum_{e\in \mathcal{E}| \mathrm{src}(e)=v}\ER_{i}(e)-\sum_{e\in \mathcal{E}| \mathrm{dst}(e)=v}\EL_{i}(e).
  \end{equation}
Here, $\mathrm{dst}(e)$ and $\mathrm{src}(e)$ are defined above in eq.~\eqref{eq:closed_path}, which take the head and tail vertices from an edge $e$.
For example, for a vertex $v$ connecting edges $a,b,c$ represented by
\begin{equation}
  \parbox{1.cm}{\includegraphics[scale=0.3]{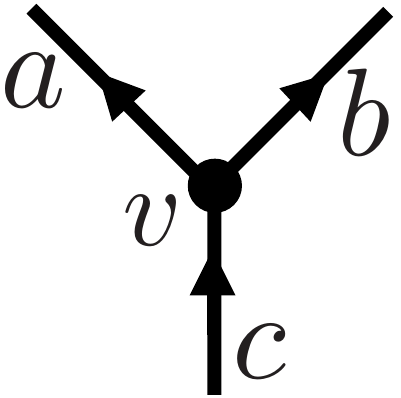}}\qquad,
  \label{eq:vabc}
\end{equation}
the Gauss-law constraint is $G_i(v)\ket{\Psi} =( R_i(a)+R_i(b)-L_i(c))\ket{\Psi}=0$.

A gauge invariant state on the vertex $v$ can be constructed by using the Clebsch–Gordan coefficients $\braket{j_an_a\, j_bn_b}{j_c,m_c}$ as
\begin{equation}
  \ket{j_aj_bj_c}=\sum_{n_a,n_b,m_c}\frac{e^{\ri\theta(j_a,j_b,j_c)}}{\sqrt{d_c}}\braket{j_an_a\, j_bn_b}{j_c,m_c}\ket{j_a,m_a,n_a}\ket{j_b,m_b,n_b}\ket{j_c,m_c,n_c},
  \label{eq:|abc>}
\end{equation}
where $1/\sqrt{d_c}$ is the normalization factor with  $d_c=2j_c+1$ such that $
\braket{j_aj_bj_c}=1$.
We can directly check $\ket{j_aj_bj_c}$ satisfies eq.~\eqref{eq:physical_state_condition} by applying $G_i(v)$.
$\theta(j_a,j_b,j_c)$ is the phase factor that cannot be determined from the normalization. We choose $e^{\ri\theta(j_a,j_b,j_c)}=(-1)^{-j_a-j_b}$, which corresponds to a phase where we no longer have to worry about the distinction between representation and anti-representation arrows.
Note that this is the property of $\mathrm{SU}(2)$, where the representation and the anti-representation are isomorphic. In general groups such as $\mathrm{SU}(3)$, representation and anti-representation should be distinct.

A general physical state is constructed by applying the Clebsch–Gordan coefficients on all vertices.
The resultant basis of states is labeled by only $\{j_e\}_{e\in\mathcal{E}}$, which is called the spin network basis used in the context of loop quantum gravity and also lattice gauge theory~\cite{penrose1971angular,Rovelli:1995ac,Baez:1994hx,Burgio:1999tg}.
Each edge can take not any $j$, but it needs to satisfy the triangular inequality at the vertex, i.e., for eq.~\eqref{eq:|abc>},
the state $\ket{j_aj_bj_c}$ needs to satisfy $j_a+j_b\geq j_c$, $j_b+j_c\geq j_a$, and $j_c+j_a\geq j_b$; otherwise, the Clebsch-Gordan coefficient vanishes.

In the following, to simplify notation, we use $a,b,c,\dots$ as labels of representation; these correspond to $j_{a}$, $j_{b}$, $j_{c}$ and etc.
It is useful to represent the state $\{j_e\}_{e\in\mathcal{E}}$ graphically; e.g., we use the following representation:
\begin{equation}
  \parbox{2.5cm}{\includegraphics[scale=0.25]{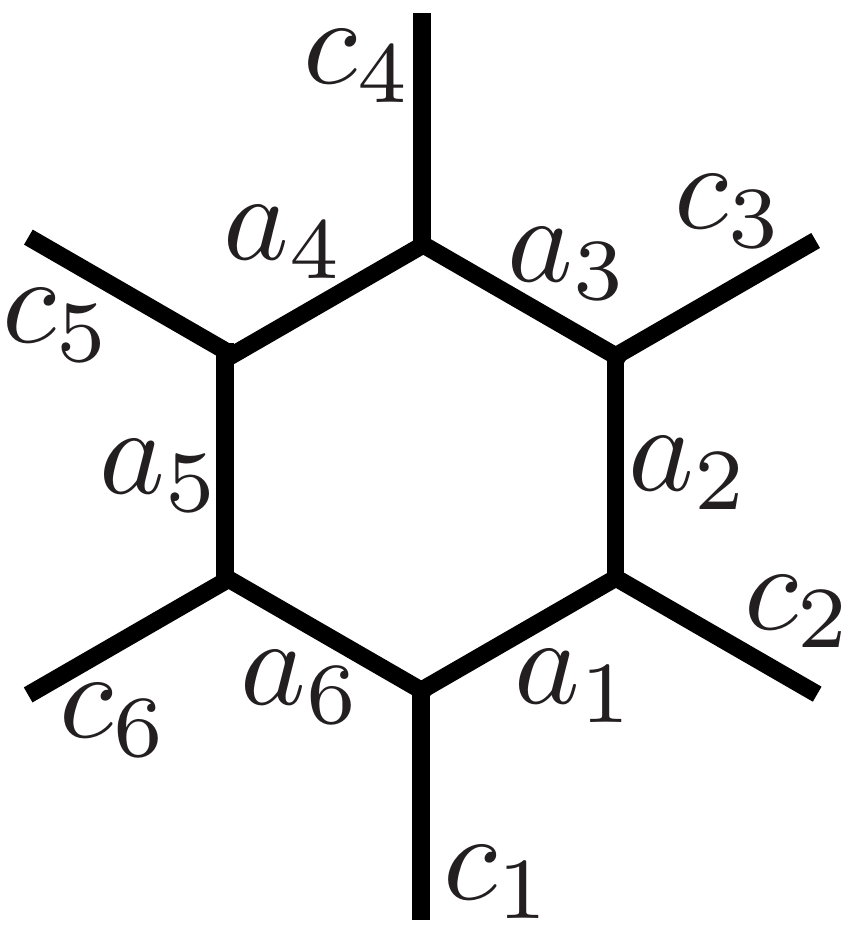}}\quad.
\end{equation}
We will use vertex labels and representation labels in a similar manner, as long as it does not cause confusion.

Let us consider the action of the Hamiltonian on the spin-network basis~\cite{Robson:1981ws}.
The action of the electric field term in the Hamiltonian, which is given by eq.~\eqref{eq:Risq} is graphically represented as
\begin{equation}
  E_i^2 \parbox{.8cm}{\includegraphics[scale=0.3]{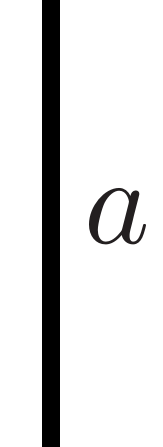}}
  = C_2(j_a)
  \parbox{.8cm}{\includegraphics[scale=0.3]{a.pdf}}.
  \label{eq:action_electric_field}
\end{equation}
On the other hand, the action of $\tr U$ is given as
\begin{equation}
  \tr U \quad
       \parbox{2.5cm}{\includegraphics[scale=0.25]{plaquetteAction0.pdf}}
       =\prod_{i=1}^{6} \sum_{a'_{i}}
       [F_{a'_{i}}^{c_{i}a_{i-1}\frac{1}{2}}]_{a_{i}{a'}_{i-1}}
         \parbox{2.5cm}{\includegraphics[scale=0.25]{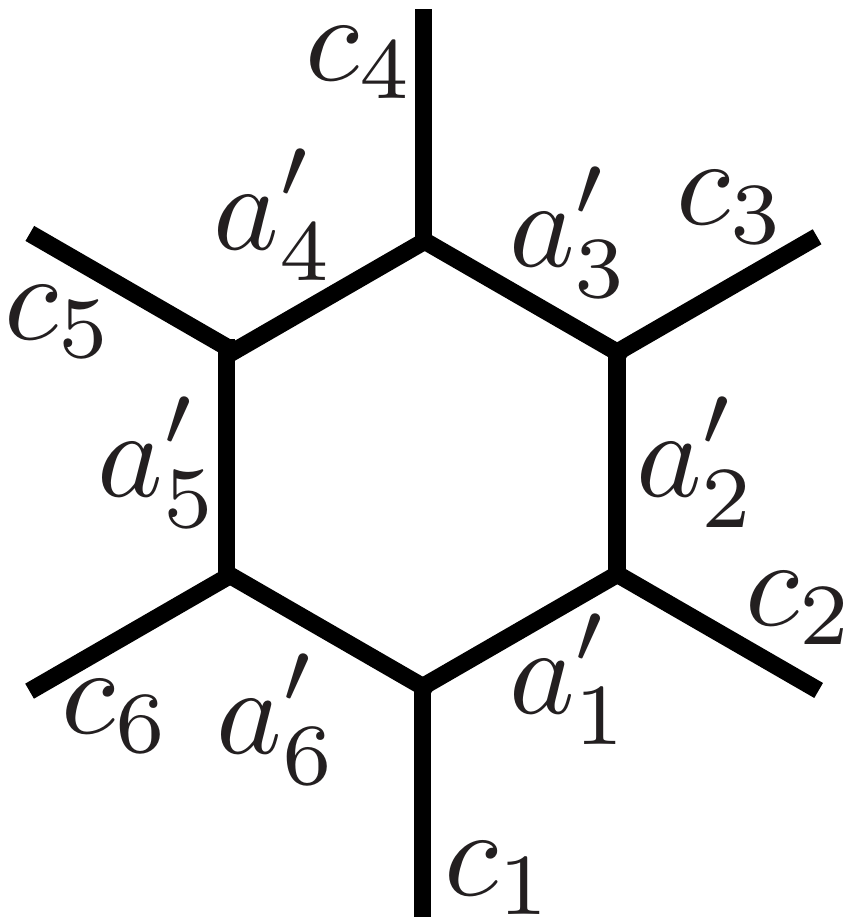}},
         \label{eq:action_trU}
\end{equation}
where $a'_0=a'_6$, and $[F^{abc}_d]_{ef}$ is the $F$-symbol given by
\begin{equation}
  \begin{split}
    &    [F^{abc}_d]_{ef}\\
&=  \sum_{m_{a},m_b,m_c,m_{e} m_{f}}
\braket{j_am_a\, j_bm_b}{j_{e}m_{e}}
\braket{j_{e}m_{e}\, j_{c}m_{c}}{j_{d}m_{d}}
\braket{j_{f}m_{f}}{ j_bm_b\, j_{c}m_{c} }
\braket{j_dm_d}{j_am_a\, j_{f}m_{f}}.
\label{eq:F}
    \end{split}
\end{equation}
Note that the right side in eq.~\eqref{eq:F} contains $m_d$, but $[F^{abc}_d]_{ef}$ does not depend on it. The $F$-symbols satisfy the so-called pentagon relation,
\begin{equation}
  [F_e^{fcd}]_{gl} [F_e^{abl}]_{fk}
  = \sum_{h} [F^{abc}_g]_{fh} [F_{e}^{ahd}]_{gk} [F^{bcd}_{k}]_{hl}.
  \label{eq:pentagon}
\end{equation}
$[F^{abc}_d]_{ef}$ can be compactly expressed using the Wigner $6\mathchar`-j$ symbol $\begin{Bsmallmatrix}
  j_a & j_b & j_e\\
  j_c & j_d & j_f
\end{Bsmallmatrix}$ as
\begin{equation}
  [F^{abc}_d]_{ef}=(-1)^{j_a+j_b+j_c+j_d}\sqrt{d_{e}d_{f}}
  \begin{Bmatrix}
    j_a & j_b & j_e\\
    j_c & j_d & j_f
  \end{Bmatrix}.
  \label{eq:F-symbol}
\end{equation}

\subsection{Multivalent graph}\label{sec:square_lattice}
We have formulated $\mathrm{SU}(2)$ gauge theory on trivalent graphs.
However, we are often interested in multivalent graphs such as square or hypercubic lattice.
We here discuss how to deal with such a case using a square lattice as an example:
\begin{equation}
  \parbox{2.3cm}{\includegraphics[scale=0.3]{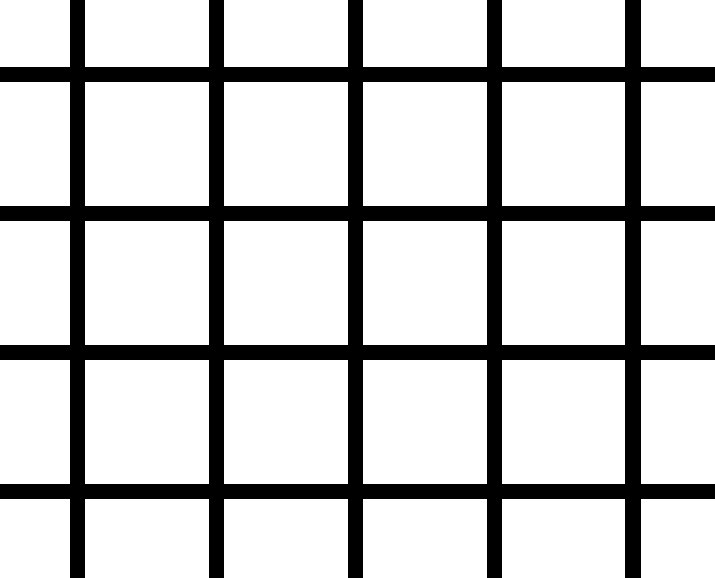}}\quad.
  \label{eq:square_lattice}
\end{equation}
The vertices are tetravalent; therefore, we cannot directly apply our previous result.
In the previous section, we can construct a gauge-invariant Hilbert space by solving the Gauss-law constraints. In this case, however, labels on the edges alone are not sufficient for a physical Hilbert space basis, and labels on the vertices are also necessary. Here, we consider a physical Hilbert space on a graph with auxiliary edges as an equivalent physical Hilbert space on eq.~\eqref{eq:square_lattice}~\cite{Anishetty:2018vod,Raychowdhury:2018tfj,PhysRevD.101.114502}.
We introduce an auxiliary edge to decompose tetravalent vertex into trivalent ones:
\begin{equation}
  \parbox{2.3cm}{\includegraphics[scale=0.3]{square.pdf}}
  \Rightarrow
  \parbox{2.3cm}{\includegraphics[scale=0.25]{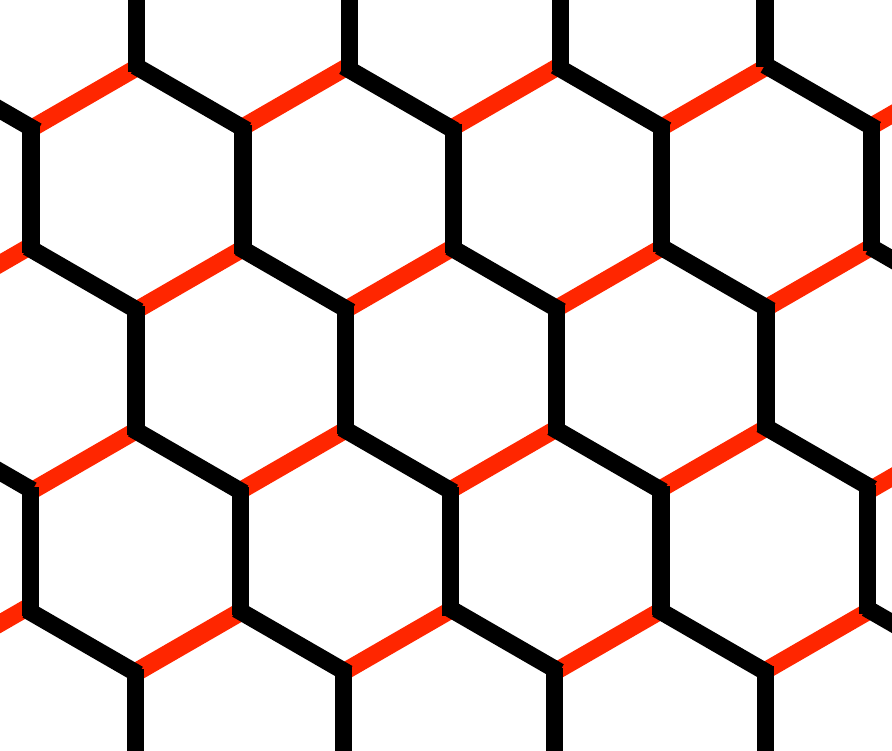}}\quad,
  \label{eq:deformation}
\end{equation}
where the red lines correspond to the auxiliary edges.
The Hilbert space of the right graph in eq.~\eqref{eq:deformation} is isomorphic to that of the left graph.
We can consider the gauge theory on this graph.
The electric fields in the Hamiltonian are assumed to act only on the black-colored edges, i.e., ${\mathcal{E}}$ is chosen as the set of black-colored edges.
On the other hand, we use the Wilson loop on the set of all minimal hexagons in the Hamiltonian.

There is a choice of auxiliary edges.
For example, we can choose the following auxiliary edges:
\begin{equation}
  \parbox{2.3cm}{\includegraphics[scale=0.3]{square.pdf}}
  \Rightarrow
  \parbox{2.3cm}{\includegraphics[scale=0.25]{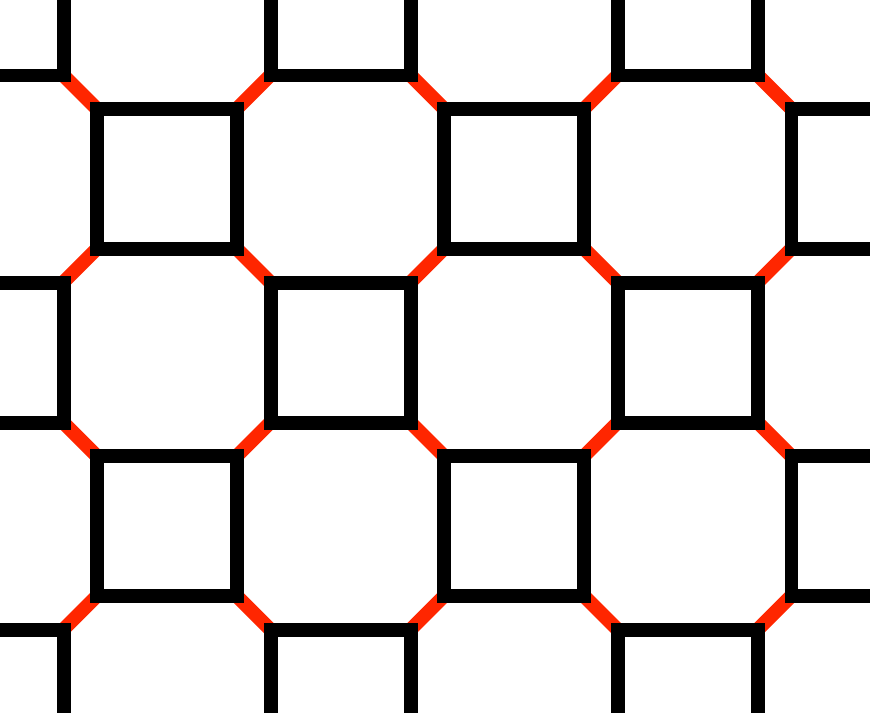}}\quad.
\end{equation}
Graphs with different auxiliary edges are related to unitary transformation by $F$-symbols, and theories with different graphs are equivalent as long as no electric field acts on the auxiliary edges.

\subsection{String net formulation and quantum deformation}\label{quantum_deformation}
The labels of the representation $j$ have no upper bound, so the dimension of the Hilbert space is infinite even on finite graphs.
In order to perform numerical calculations, it is necessary to regularize the Hilbert space to a finite dimension.
A naive cutoff may no longer guarantee certain properties of the $F$-symbols, such as eq.~\eqref{eq:pentagon}, or equivalently the independence of the choice of auxiliary edges discussed in the previous subsection.
In other words, there appear infinitely many models due to choice of auxiliary edges, and we need to confirm that they are converged to the same theory in the large cutoff limit.
Quantum deformation ($q$-deformation) of gauge groups provides a regularization maintaining those properties of the $F$ symbols manifestly, so that graphs or regularized models with different auxiliary edges are unitary equivalent at finite cutoff.
Here we consider $\mathrm{SU}(2)_k$ gauge theory as a regularized model discussed in refs.~\cite{Bimonte:1996fq,Burgio:1999tg,Dittrich:2018dvs}, and we give an explicit form of the Hamiltonian as a string-net model that can be implemented numerically
(See ref.~\cite{Cunningham:2020uco} for a regularization based on the $q$-deformation in the path integral formulation).
The details of the quantum group are not necessary for our purpose, and only the essential facts will be presented.
If you would like to learn more about quantum groups and $\mathrm{SU}(2)_k$, see, e.g., refs.~\cite{KauffmanLins+1994,CarterFlathSaito+1996,Bonatsos:1999xj}.

In $\mathrm{SU}(2)_k$, $j$ takes $0,1/2,1,\cdots, k$, i.e., the number of states is $k+1$. This corresponds to the cutoff $j_{\mathrm{max}}=k/2$.
Correspondingly, $ j_a+j_b+j_c\leq k$ is added to the triangle relation,  $ j_{a}+j_{b}\geq j_{c}, j_{b}+j_{c}\geq j_{a}, j_{c}+j_{a}\geq j_{b}, j_{a}+j_{b}+j_{c}\in\mathbb{Z}$.
Roughly speaking, in $\mathrm{SU}(2)_k$, an integer $n$ is replaced by a $q$-number $[n]$, where 
$[n]$ is defined as
\begin{equation}
  [n]\coloneqq  \frac{q^{\frac{n}{2}}-q^{-\frac{n}{2}}}{q^{\frac{1}{2}}-q^{-\frac{1}{2}}}=\frac{\sin\qty(\frac{\pi}{k+2}n)}{\sin\qty(\frac{\pi}{k+2})},
  \label{eq:[n]}
\end{equation}
with $  q = \exp\ri \frac{2\pi}{k+2}$.
For example, the dimension of representation $j_a$ becomes $d_a=[2j_a+1]$.
Similarly, the second Casimir invariant of representation $j_a$ is 
\begin{equation}
  C_2(j_a)=[j_a][j_a+1].\label{eq:C2q}
\end{equation}
The $F$-symbols can be represented as the same form in eq.~\eqref{eq:F-symbol},
\begin{equation}
  [F^{abc}_d]_{ef}=(-1)^{j_a+j_b+j_c+j_d}\sqrt{d_{e}d_{f}}
  \begin{Bmatrix}
    j_a & j_b & j_e\\
    j_c & j_d & j_f
  \end{Bmatrix},
  \label{eq:F_symbol_q}
\end{equation}
where the $q$-deformed $6\mathchar`-j$ symbol is 
\begin{equation}
  \begin{split}
    \begin{Bmatrix}
      a & b & c\\
      d & e & f
    \end{Bmatrix}
    \coloneqq \Delta(a,b,c)\Delta(a,e,f)\Delta(d,b,f)\Delta(d,e,c)\sum_z(-1)^z [z+1]!\\   
    \times \frac{([a+b+d+e-z]![a+d+c+f-z]![b+e+c+f-z]!)^{-1}}{[z-a-b-c]![z-a-e-f]![z-d-b-f]![z-d-e-c]!},
  \end{split}
\end{equation}
where
\begin{equation}
  \Delta(a,b,c)=\delta_{abc}\sqrt{\frac{[a+b-c]![a-b+c]![-a+b+c]!}{[a+b+c+1]!}},
\end{equation}
and $[n]!\coloneqq[n][n-1]\cdots [2][1]$.
Here, $z$ satisfies
\begin{equation}
  \max(a+b+c,a+e+f,d+b+f,d+e+c)\leq z\leq 
  \min(a+b+d+e,a+d+c+f,b+e+c+f).
\end{equation}
These values are reduced to the $\mathrm{SU}(2)$ ones, by replacing $[n]$ with $n$ for an integer $n$.
The Hamiltonian of $\mathrm{SU}(2)_k$ and its action are the same form as in eqs.~\eqref{eq:Hamiltonian}, ~\eqref{eq:action_electric_field} and \eqref{eq:action_trU}.

It is convenient to represent this system in a type of spin model called a string-net model~\cite{Levin:2004mi}.
Since there are $(k+1)$ degrees of freedom on each edge, this can be regarded as a $S=k/2$ spin system.
Note that the vertices are constrained to satisfy the triangle inequality. In a spin model, those constraints can be realized by adding a penalty term into the Hamiltonian.
For this purpose, we introduce the following function,
\begin{equation}
 \delta_{abc}\coloneqq \begin{cases}
  1 &
  j_{a}+j_{b}\geq j_{c},
  j_{b}+j_{c}\geq j_{a},
  j_{c}+j_{a}\geq j_{b},
  j_{a}+j_{b}+j_{c}\in\mathbb{Z},\
   \text{and}\ j_a+j_b+j_c\leq k
  \\
  0 & \text{else}
  \end{cases}.
  \label{eq:fusion q}
  \end{equation}
$\delta_{abc}=1$ if $a$, $b$, and $c$ satisfy the triangular inequality; otherwise it is zero.
Using this function, we define the penalty term $Q(v)$
on a vertex $v$ connecting to edges $a,b,c$, 
whose action on a state is
\begin{equation}
  Q(v) \ket{j_aj_bj_c}=  \delta_{abc}\ket{j_aj_bj_c}.
\end{equation}
Note that $Q(v)$ commutes with $E_i^2(e)$ and $\tr U(f)$, like the generators of a gauge transformation.
Although $Q(v)$ does not generate a gauge transformation, we will also refer to it as a Gauss-law constraint in the following because it constrains the Hilbert space of the spin model to a gauge-invariant subspace.

In summary, the $\mathrm{SU}(2)_k$ string-net model of lattice Yang-Mills theory is a spin model with $S=k/2$ spins on edges. The Hamiltonian consists of three parts: electric, magnetic, and penalty terms:
\begin{equation}\label{eq:HYM}
  H_\mathrm{YM} = 
  \frac{1}{2}\sum_{e\in {\mathcal{E}}} E_i^2(e)
  - K \sum_{f\in \mathcal{F}} \tr U (f)
  -t\sum_{v\in\mathcal{V}}Q(v).
\end{equation}
Here, $K$ and $t$ are coupling constants.
To restrict the low-energy space to the physical Hilbert space, $t$ must be sufficiently large.
The actions of operators on a state are graphically represented as
\begin{align}
  E_i^2 \parbox{.8cm}{\includegraphics[scale=0.3]{a.pdf}}
  &= C_2(j_a)
  \parbox{.8cm}{\includegraphics[scale=0.3]{a.pdf}},
  \label{eq:action_electric_field_q}
  \\
  \tr U \quad
  \parbox{2.5cm}{\includegraphics[scale=0.25]{plaquetteAction0.pdf}}
  &=\prod_{i=1}^{6} \sum_{a'_{i}}
  [F_{a'_{i}}^{c_{i}a_{i-1}\frac{1}{2}}]_{a_{i}{a'}_{i-1}}
    \parbox{2.5cm}{\includegraphics[scale=0.25]{plaquetteAction2.pdf}},\\
  Q \  \parbox{1.5cm}{\includegraphics[scale=0.3]{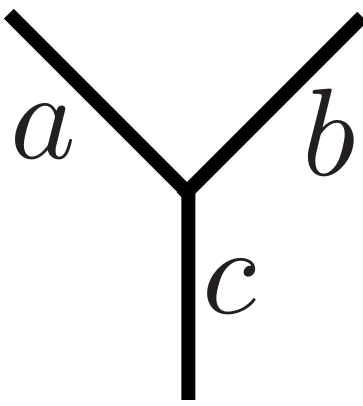}}&=  \delta_{abc}\  \parbox{1.5cm}{\includegraphics[scale=0.3]{abc.pdf}},
  \label{eq:action_trU_q}
\end{align}
where $C_2(j_a)$, $[F^{abc}_d]_{ef}$, and $\delta_{abc}$ are given in eqs.~\eqref{eq:C2q}, \eqref{eq:F_symbol_q}, and \eqref{eq:fusion q}, respectively.
Note that the same procedure can be extended to $\mathrm{SU}(3)_k$. However, since $\mathrm{SU}(3)_k$ has multiplicity, vertex labels as well as edge labels, are required to specify the physical state (see ref.~\cite{Hayata:2023bgh}).
Cutoff dependence of the $q$-deformation in a single plaquette model for $\mathrm{SU}(2)_k$ and $\mathrm{SU}(3)_k$ is discussed in appendix~\ref{sec:appendix}.
From the single plaquette model, we estimated that $k$ is required to be $\mathcal{O}(10)$ for simulating the groundstate in the large $k$ limit with $K\sim 1$.

If the electric part is dropped and the Wilson loop with the fundamental representation is replaced by one with the regular representation, i.e.,
\begin{equation}
  \tr U (f) \to \tr U_\mathrm{R} (f) =\sum_{a}d_a\tr U_a (f),
\end{equation}
the model reduces to
\begin{equation}
  H_\mathrm{LW} =
    - \frac{K}{\mathcal{D}^2}\sum_{a} d_a\tr U_a (f)
    -t\sum_{v\in\mathcal{V}}Q(v).
    \label{eq:HLW}
\end{equation}
Here, $U_a (f)$ represents the Wilson loop with the representation $a$.
We rescaled $K$ to $K/\mathcal{D}^2$, and $\mathcal{D}$ is the total quantum dimension defined by
\begin{equation}
  \mathcal{D}^2 \coloneqq \sum_{j_a=0}^{k/2}d_a^2=\frac{k+2}{2}\frac{1}{\sin^2\qty(\frac{\pi}{k+2})}.
\end{equation}
This model (with $t=1$ and $K=1$) is known as the Levin-Wen model, whose ground state exhibits topological order with non-Abelian anyons~\cite{Levin:2004mi}.
To compare it with the string-net model of lattice Yang-Mills theory, we study the spectrum of  the perturbed Levin-Wen model by adding the electric term~\cite{Schulz:2012em,Schulz:2014mta,Dusuel:2015sta,Schotte:2019cdg},
\begin{equation}
  H_\mathrm{pLW} =
  \frac{1}{2}\sum_{e\in {\mathcal{E}}} E_i^2(e)
    - \frac{K}{\mathcal{D}^2}\sum_{a} d_a\tr U_a (f)
    -t\sum_{v\in\mathcal{V}}Q(v),
    \label{eq:HLW_perturbed}
\end{equation}
in the next section.

\section{Exact diagonalization}
\label{sec:ED}

As an application of the string-net basis representation of the $\mathrm{SU}(2)_k$ Hamiltonian~\eqref{eq:HYM}, we compute its spectrum using exact diagonalization and discuss the so-called quantum many-body scars~\cite{bernien_probing_2017,turner_weak_2018} in section~\ref{sec:YM}. We call the model~\eqref{eq:HYM} the Yang-Mills model.
It is known that the constrained Hilbert space due to the Gauss laws plays an essential role in quantum many-body scars~\cite{Banerjee:2020tgz,Biswas:2022env}, and thus it is interesting to study whether the nonabelian Gauss law can host them.

To further understand this phenomenon, we ask whether they need only the Gauss-law constraints.
In section~\ref{sec:LW},  we study the perturbed Levin-Wen model~\eqref{eq:HLW_perturbed}, which shares the same Gauss-law constraints with the Yang-Mills model, but has a different representation of the Wilson loop operator.

\subsection{Yang-Mills model}\label{sec:YM}

\begin{figure}[t]
\begin{center}
 \includegraphics[width=.5\textwidth]{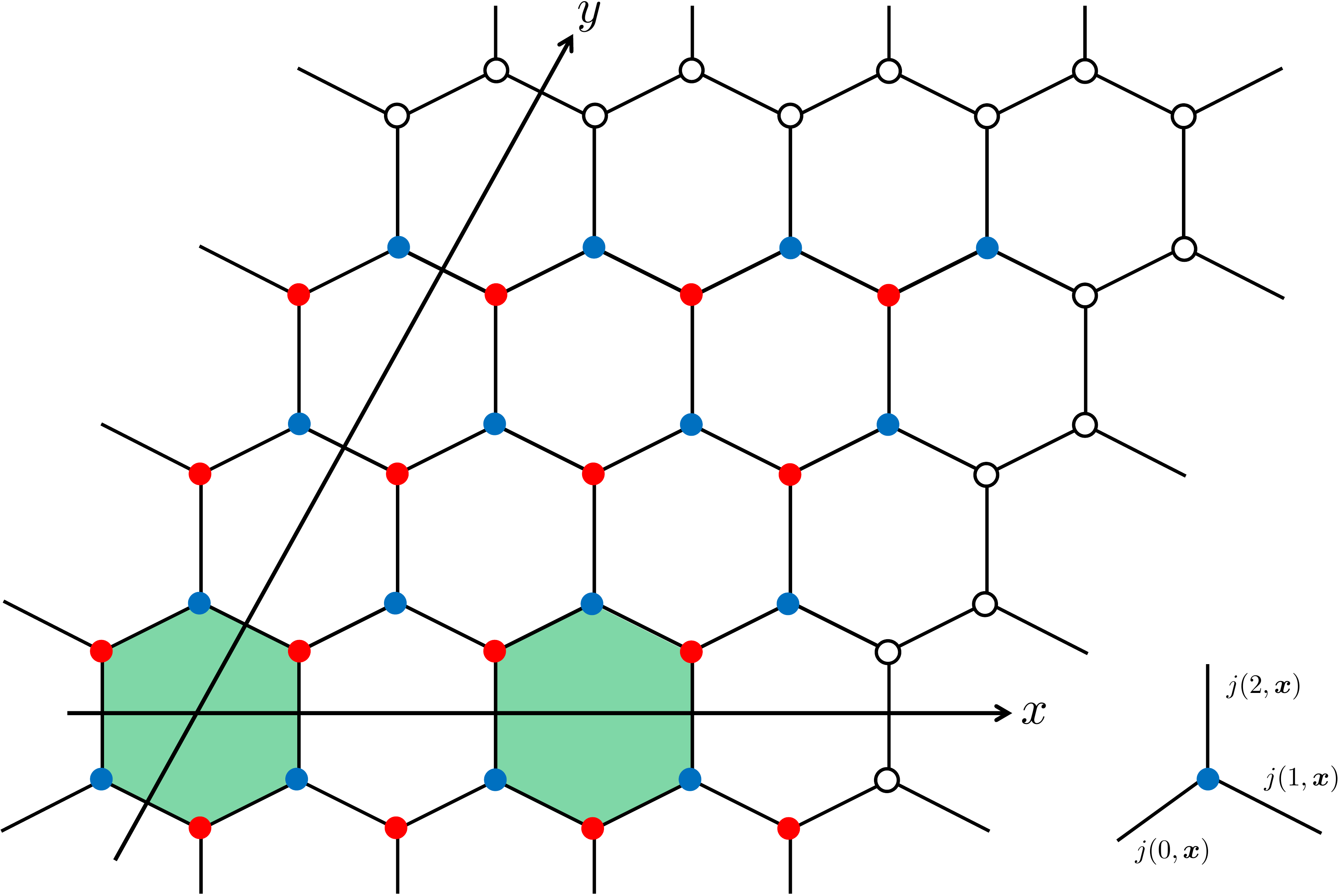}
\caption{\label{fig:latticegeometry}Lattice geometry with $N_x=N_y=4$. Link variables in the unit cells are grouped, and labeled as $j(a,\bm x)$ with $\bm x$ being the position of the unit cell. 
The blue dots show the center of the unit cells. The Gauss law is imposed at blue and red vertices. The periodic boundary conditions are imposed at open circles.}
\end{center}
\end{figure}

\begin{table*}[htb]
\begin{center}
  \begin{tabular}{c|c|ccccc}
   Model &($N_x,N_y,j_\mathrm{max}$) & Total & Gauss law &  Winding  &  Scars   \\
    \hline
    \multirow{2}{*}{Yang-Mills}&$(4,4,\frac{1}{2})$ & $2^{48}$ & $131072$ & $32768$ & $3$   \\
     &$(4,2,1)$ & $3^{24}$ & $131328$ & $33024$ & $29$   \\
    \hline
    \multirow{2}{*}{Levin-Wen}&$(4,2,\frac{3}{2})$ & $2^{24}$ & $29375$ & - & $0$ \\
     &$(4,2,1)$ & $3^{24}$ & $131328$ & $33024$ & $3$ \\
  \end{tabular}
  \caption{\label{table:dim}The number of states in the full Hilbert space, in the Hilbert space satisfying the Gauss-law constraints (i.e., $Q(v)=1$ sector) [see ref.~\cite{Vidal:2021isf} for a general expression of the Hilbert space dimension], in the Hilbert space satisfying the Gauss-law constraints and with zero winding numbers, and in scars. The upper two rows represent the results of the Yang-Mills theory, while the lower two rows are those of the perturbed Levin-Wen model. In the Levin-Wen model with $j_\mathrm{max}=3/2$, $j$ is restricted to integers.}
\end{center}
 \end{table*}
\begin{figure}[t]
\begin{center}
 \includegraphics[width=.48\textwidth]{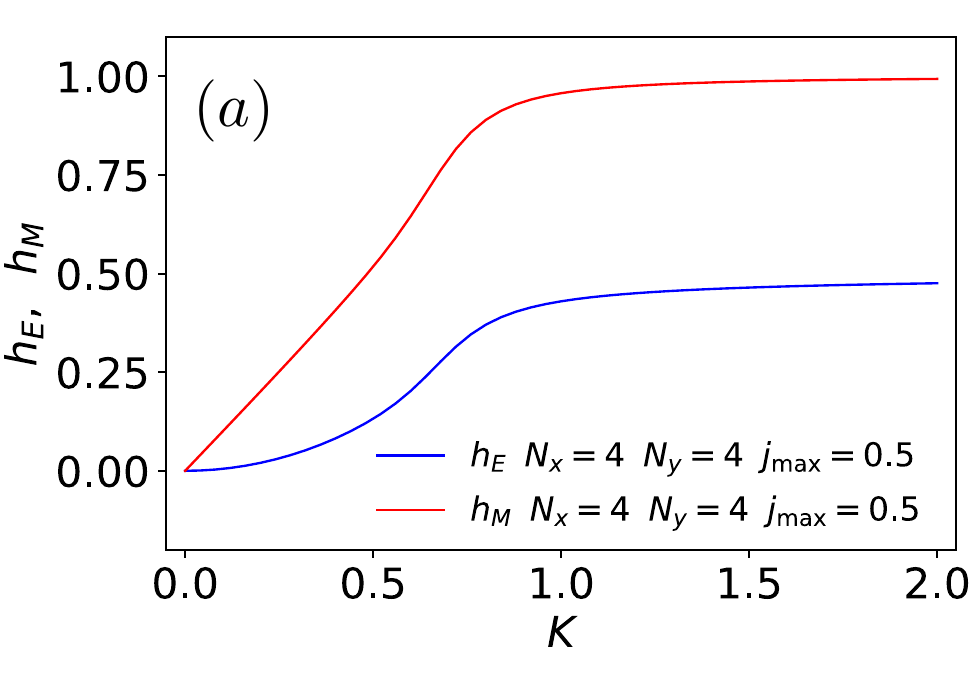}
 \includegraphics[width=.47\textwidth]{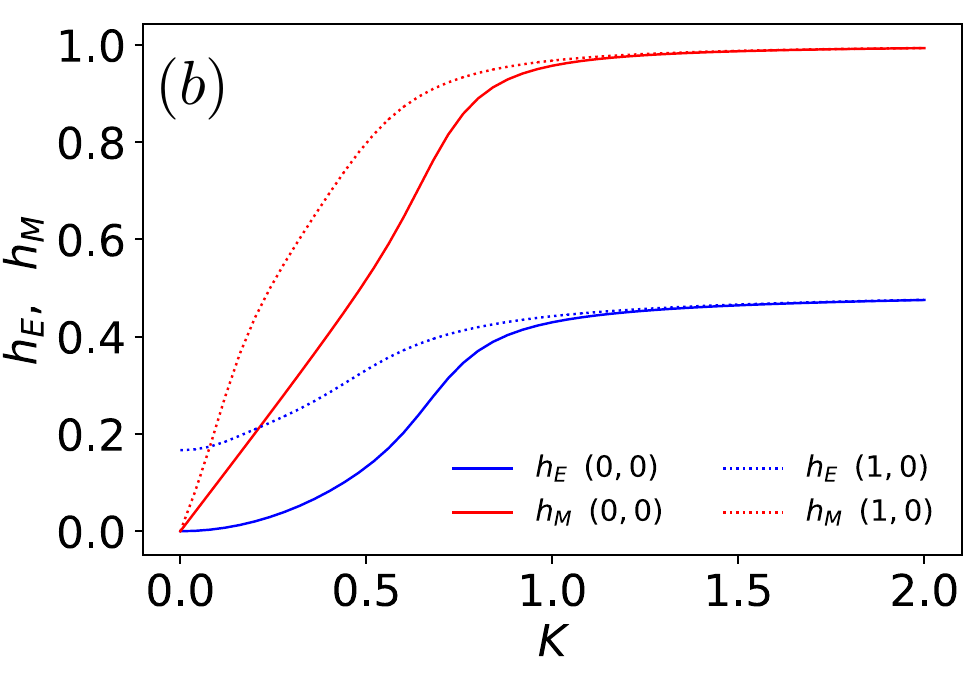}
\caption{\label{fig:vev}(a) Groundstate expectation values of electric and magnetic Hamiltonians as a function of coupling strength $K$. 
(b) Comparison with different winding sectors. Three sectors ($1,0$), ($0,1$), and ($1,1$) are degenerate, so that only the ($1,0$) sector is shown in the figure.
}
\end{center}
\begin{center}
 \includegraphics[width=.46\textwidth]{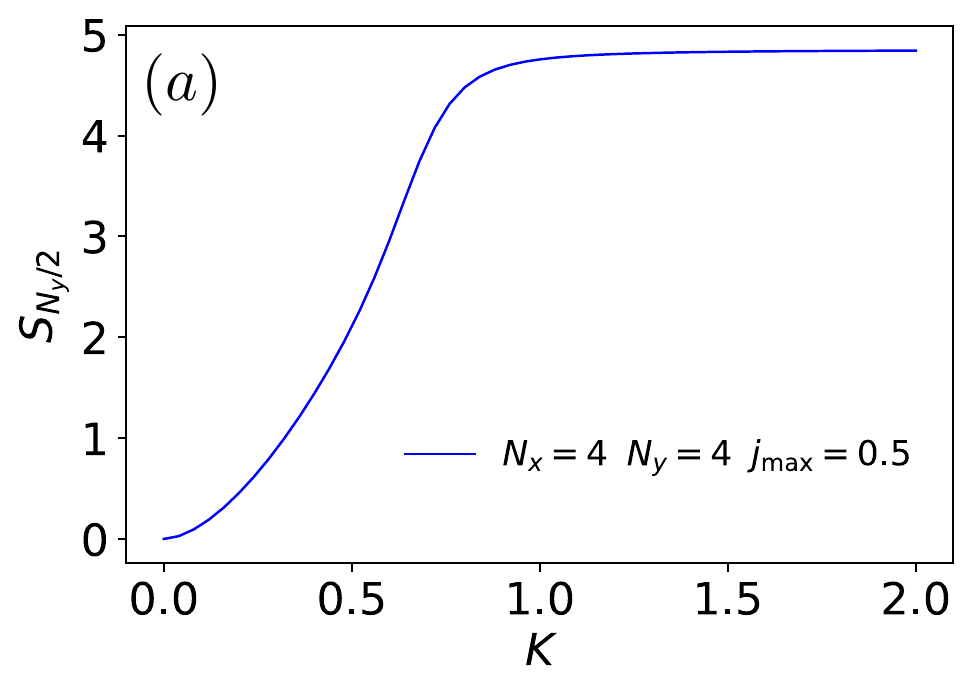}
  \includegraphics[width=.48\textwidth]{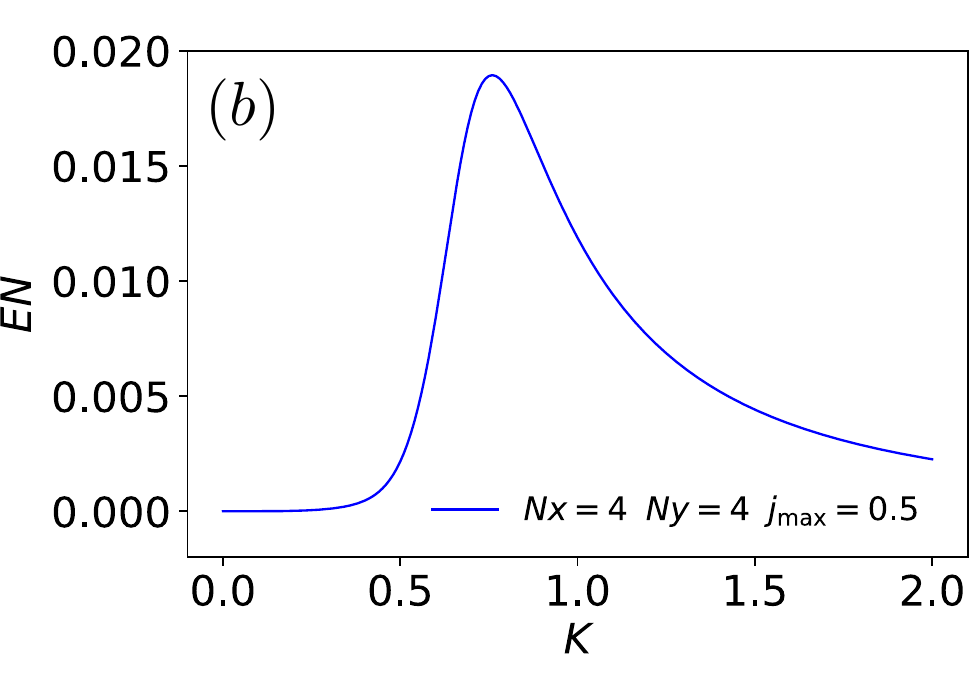}
\caption{\label{fig:vevEE}(a)Bipartite entanglement entropy of the groundstate as a function of the coupling strength $K$.
(b)Logarithmic entanglement negativity of the groundstate as a function of the coupling strength $K$.}
\end{center}
\end{figure}

We consider the Yang-Mills model~\eqref{eq:HYM} on a honeycomb lattice shown in figure~\ref{fig:latticegeometry}. We impose the periodic boundary conditions along the $x$ and $y$ directions, and $N_x$ and $N_y$ are the lengths of those directions. The total number of the unit cells (plaquettes) is $V=N_xN_y$, and that of edges is $3N_xN_y$.
Link variables living on edges are labeled by $j(a,\bm x)$, where $a$, and $\bm x=(x,y)$ are internal indices in the unit cell, and two-dimensional positions of the unit cells, respectively (see figure~\ref{fig:latticegeometry}).
In the periodic boundary conditions, the parity of $2j(a,\bm x)$, which intersect with the $x$ or $y$ axis, i.e., $\sum_y 2j(1,0,y)$ [mod $2$] and $\sum_x 2j(2,x,0)$ [mod $2$] are conserved (see figure~\ref{fig:latticegeometry}).
This corresponds to $\mathbb{Z}_2$ $1$-form symmetry~\cite{Gaiotto:2014kfa}, which divides the Hilbert space into distinct topological sectors characterized by the binary winding numbers ($W_x,W_y$)=($0,0$), ($1,0$), ($0,1$), and ($1,1$). We focus on the largest sector with ($W_x,W_y$)=($0,0$).

First, we consider $N_x=4$, $N_y=4$, and $j_\mathrm{max}=1/2$, i.e., $k=1$.
From the total Hilbert space, we pick up the subspace by imposing $Q(v)=1$, $W_x=0$, and $W_y=0$, and we diagonalize the Hamiltonian in the subspace.
The dimension of the Hilbert space is summarized in table~\ref{table:dim}.
Before discussing quantum scars, let us see the properties of the groundstate $\ket{\Omega}$ in this model.
In figure~\ref{fig:vev}(a), we show the groundstate expectation value of the electric and magnetic parts of the energy density,
which are given explicitly as
\begin{align}
  {h}_\mathrm{E} = \frac{1}{2V}\sum_{e\in {\mathcal{E}}} \bra{\Omega}{E_i^2(e)}\ket{\Omega},\quad
  {h}_\mathrm{M} =  \frac{1}{V}\sum_{f\in \mathcal{F}} \bra{\Omega}\tr U (f)\ket{\Omega},
\end{align}
respectively.
We also show the bipartite entanglement entropy of the groundstate in figure~\ref{fig:vevEE}.
By partitioning the lattice into two equal parts $A$ and $\bar{A}$ along the $y$ direction, and tracing out $\bar{A}$, we compute the reduced density matrix of the $A$ subspace as 
\begin{equation}
  \rho_A = \tr_{\bar{A}} |\Psi\rangle\langle\Psi| ,
  \label{eq:densitymatrix}
\end{equation}
where $|\Psi\rangle$ is the energy eigenstate in our case.
From the reduced density matrix, we compute the bipartite entanglement entropy $S_{N_y/2}$ as
\begin{equation}
  S_{N_y/2} = - \tr_{A} \rho_A \log \rho_A .
\end{equation}
Note that $\rho_A=\rho^\dag_A$ as is evident from eq.~\eqref{eq:densitymatrix} and thus $S_{N_y/2}$ can be easily computed from the eigenvalues $e_A$ of $\rho_A$ as $S_{N_y/2} = -\sum_A  e_A \log e_A$.
Although the spatial size of the lattice is so limited, and thus we cannot judge the order of the phase transition, we can see two distinct phases in the weakly and strongly coupling regimes.
At the strong coupling regime (i.e., small $K$), the expectation value of $H_\mathrm{M}$ is small; that is, the system is in the confinement phase, where the bipartite entanglement entropy is also small. In the strong coupling limit $(K=0)$, the vacuum state is the product state with $j(a,\bm{x})=0$, so that entanglement entropy vanishes.
On the other hand, at the weak coupling regime (i.e., large $K$), the expectation value of $H_\mathrm{M}$ is large; that is, the system is in the toplogical phase, where the bipartite entanglement entropy is also large.
The toplogical phase corresponds to the broken phase of $\mathbb{Z}_2$ $1$-form symmetry.
Reflecting this, the groundstate energies of the different topological sectors are degenerate.
Figure~\ref{fig:vev}(b) shows the degeneracy of ${H}_\mathrm{E}$ and ${H}_\mathrm{M}$ with different winding sectors at a weak coupling region.
These observations indicate that a phase transition occurs between the weakly and strongly coupled regions in the thermodynamic limit.
To extract information of the phase transition point from limited data, we further compute the logarithmic entanglement negativity~\cite{Vidal:2002zz,Plenio:2005cwa,2006PhRvL..97q0401D}. 
By partitioning the system into three regions $A$, $B$, and $C$, and tracing out $C$, we compute the reduced density matrix of the $AB$ subspace $\rho_{AB}$ as 
\begin{equation}
  \rho_{AB} = \tr_{C} |\Psi\rangle\langle\Psi| =\sum_{A,B}\sum_{A^\prime,B^\prime}\rho_{AB,A^\prime B^\prime}|A,B\rangle\langle A^\prime, B^\prime|.
  \label{eq:densitymatrix2}
\end{equation}
Then, we define the partial transpose $\rho^{T_B}_{AB}$ as
\begin{equation}
  \rho^{T_B}_{AB} =\sum_{A,B}\sum_{A^\prime,B^\prime}\rho_{AB^\prime,A^\prime B}|A,B\rangle\langle A^\prime, B^\prime|.
  \label{eq:densitymatrix_PT}
\end{equation}
Using it, the logarithmic entanglement negativity is given as 
\begin{equation}
  EN = \log\sum_n \sigma_n(\rho^{T_B}_{AB}) ,
\end{equation}
where $\sigma_n(\rho^{T_B}_{AB})$ are the singular values of $\rho^{T_B}_{AB}$.
We show the logarithmic entanglement negativity of the groundstate in figure~\ref{fig:vevEE}(b).
As regions $A$ and $B$, we chose $j(a,\bm x)$ defined on the edges of disjoint plaquettes colored in green in figure~\ref{fig:latticegeometry}. We found that the logarithmic entanglement negativity becomes maximum at $K=0.76$.
Also, its slope is different between confined and toplogical phases, which implies that it would be singular at the phase transition point around $K=0.76$ in the thermodynamic limit.
Ref.~\cite{Zache:2023dko} studied the phase transition of $\mathrm{SU}(2)_k$ Yang-Mills theory using a mean-field type variational ansatz, where the critical $K$ increases as $k$ or $j_{\rm max}$ increases ($j_{\rm max}=k/2$). 
Using the mean-field computation, the critical $K$ of the Yang-Mills model~\eqref{eq:HYM} on a honeycomb lattice is estimated as $K=1.0$ for $j_{\rm max}=1/2$.
We also extended their mean-field computation and studied the phase transition of $\mathrm{SU}(3)_k$ Yang-Mills theory in ref.~\cite{Hayata:2023bgh}. 

\begin{figure}[t]
\begin{center}
 \includegraphics[width=.48\textwidth]{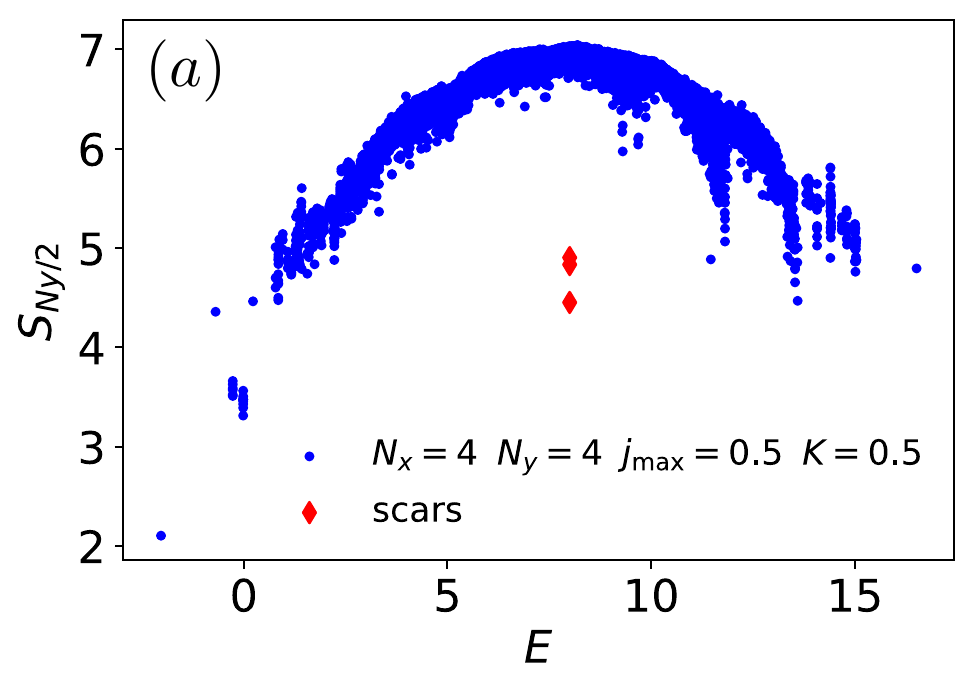}
 \includegraphics[width=.48\textwidth]{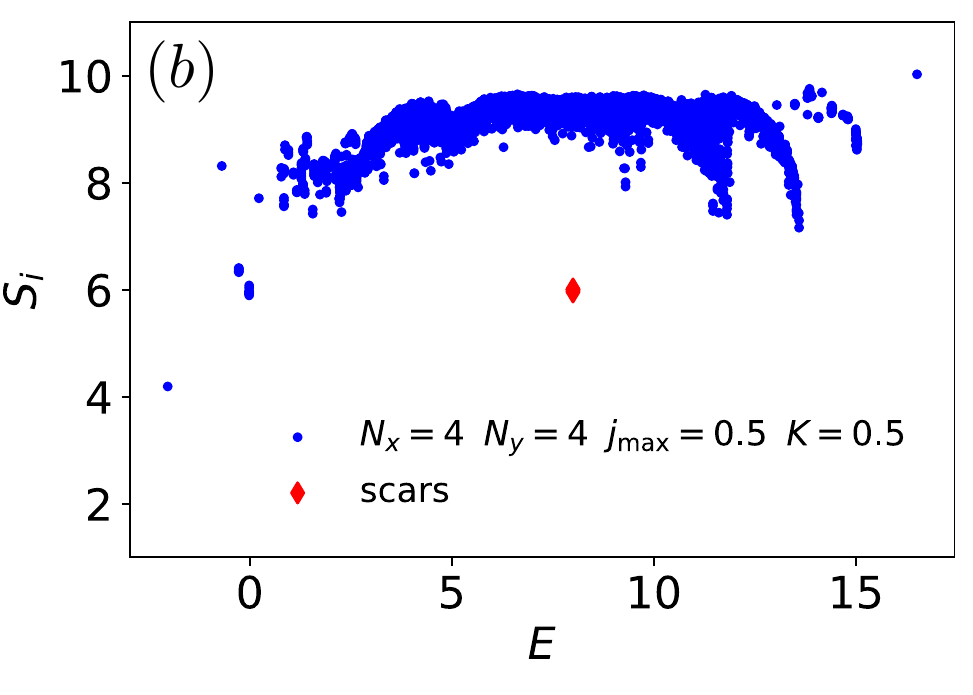}
\caption{\label{fig:EEj0.5}(a) Bipartite entanglement entropy for all eigenvectors at $j_{\rm max}=0.5$ and $K=0.5$.
(b) Shannon entropy for all eigenvectors at $j_{\rm max}=0.5$ and $K=0.5$.}
\end{center}
\end{figure}

\begin{figure}[t]
\begin{center}
 \includegraphics[width=.48\textwidth]{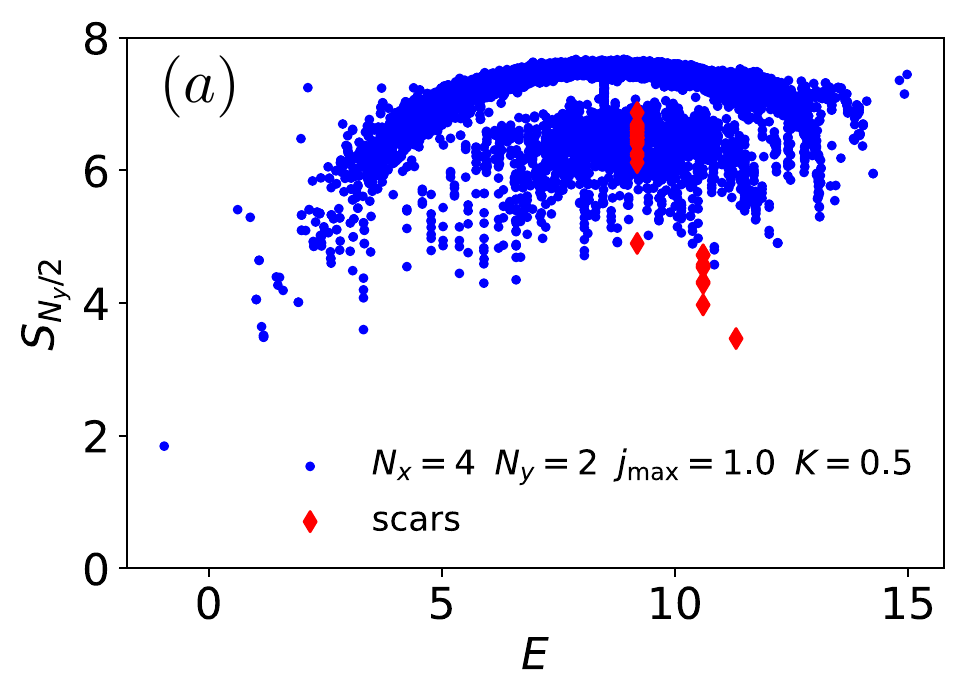}
 \includegraphics[width=.48\textwidth]{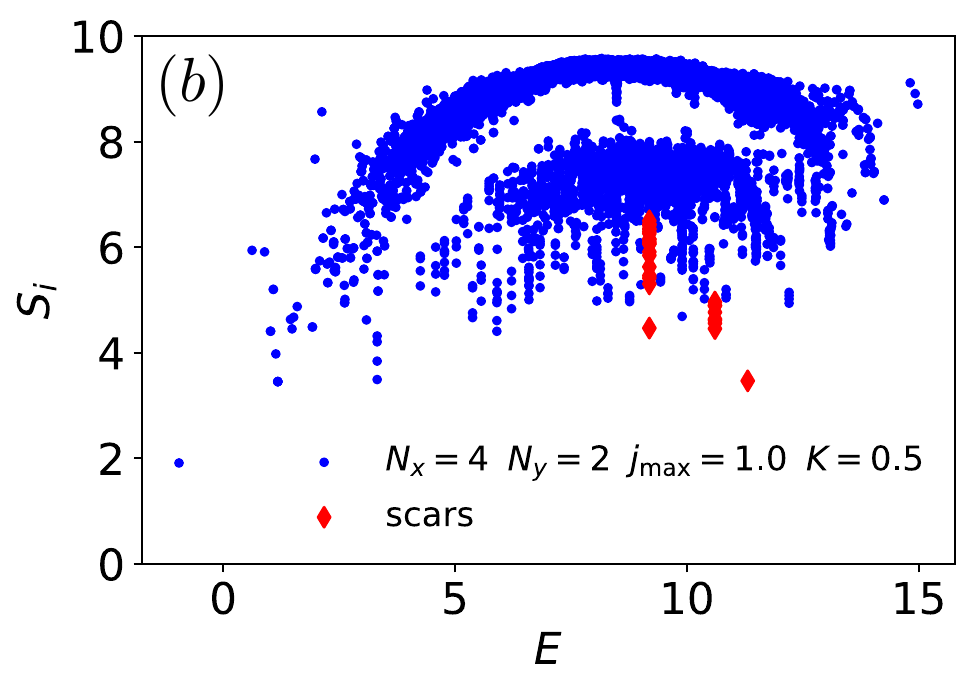}
\caption{\label{fig:EEj1.0}(a) Bipartite entanglement entropy for all eigenvectors at $j_{\rm max}=0.5$ and $K=0.5$. (b) Shannon entropy for all eigenvectors at $j_{\rm max}=0.5$ and $K=0.5$.}
\end{center}
\end{figure}

We now discuss the properties of the excited states. We compute the bipartite entanglement entropy $S_{N_y/2}$ for all energy eigenstates as well as the Shannon entropy, 
\begin{equation}
  S_i\coloneqq-\sum_n |\Psi(n)|^2\log |\Psi(n)|^2,
\end{equation}
where $\Psi(n)=\langle n|\Psi\rangle$ is the many-body wave function, and $n$ labels basis of the physical Hilbert space.
The Shannon entropy quantifies the localization of the eigenvectors in the computational basis, i.e., the string net basis since it is maximized when $|\Psi(n)|^2$ is uniformly distributed.
The results are shown in figure~\ref{fig:EEj0.5}. In the mid-spectrum, there exist only three eigenstates with low entanglement entropy.
Their entropy is substantially small compared with that of other eigenstates with the same energy, which leads to violation of the eigenstate thermalization hypothesis~\cite{PhysRevA.43.2046,Srednicki:1994mfb,2008Natur.452..854R}. As observed in the Shannon entropy, these states show the localization in the Hilbert space.
These states are referred to as quantum many-body scars from zero modes in the literature~\cite{Banerjee:2020tgz,Biswas:2022env}.
We have verified that those eigenstates are simultaneous eigenstates of the electric and magnetic Hamiltonians in particular, they have zero eigenvalue for the magnetic Hamiltonian (i.e., they are zero modes). Therefore, their wave functions and the expectation values are independent of $K$, which are important features of the scars from zero modes. Our numerical diagonalization demonstrates that quantum scars arise in a nonabelian lattice gauge theory.

Next, in order to examine the effect of truncation, we diagonalize the Hamiltonian with $N_x=4$, $N_y=2$, and $j_\mathrm{max}=1$ ($k=2$).
The dimension of the physical Hilbert space is summarized in table~\ref{table:dim}.
We show the bipartite entanglement entropy $S_{N_y/2}$ and the Shannon entropy $S_i$ in figure~\ref{fig:EEj1.0}.
We find $29$ scars, where the dimension of the physical Hilbert space is the same order as the case with $j_\mathrm{max}=1/2$, $N_x=4$, and $N_y=4$.
This suggests that by increasing $j_\mathrm{max}$, there appear to be a lot of scars.
We have verified that they are simultaneous eigenstates of the electric and magnetic Hamiltonians, and their eigenvalues are zero for the magnetic Hamiltonian. 

\subsection{Levin-Wen model}\label{sec:LW}

\begin{figure}[t]
\begin{center}
 \includegraphics[width=.47\textwidth]{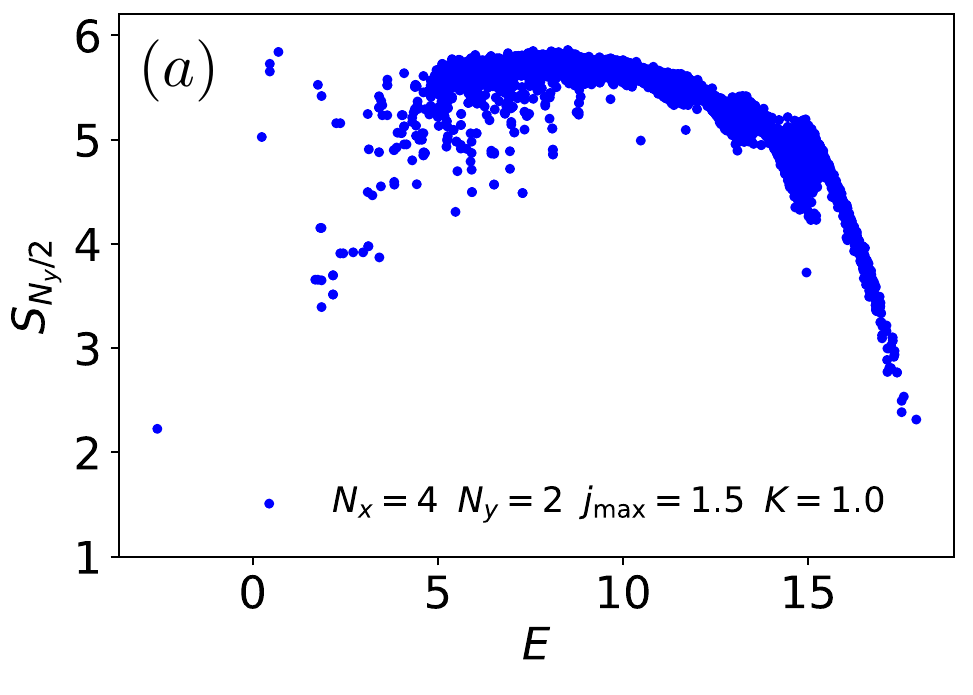}
 \includegraphics[width=.48\textwidth]{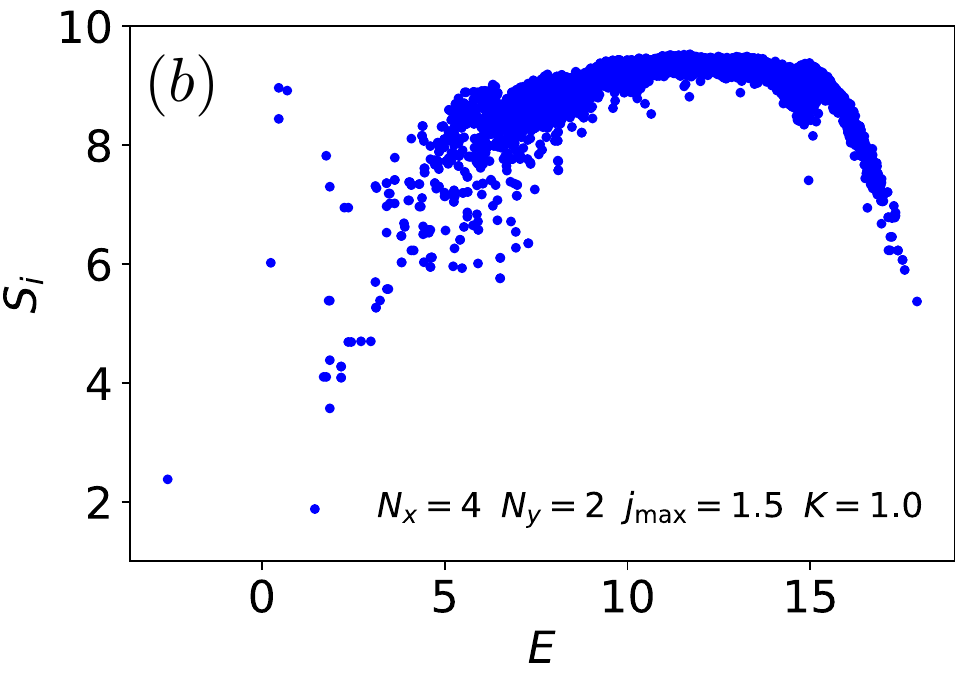}
\caption{\label{fig:EEj1.5}(a) Bipartite entanglement entropy for all eigenvectors of the perturbed Levin-Wen model at $j_{\rm max}=1.5$, $K=1.0$. (b) Shannon entropy for all eigenvectors of the perturbed Levin-Wen at $j_{\rm max}=1.5$, $K=1.0$. No quantum scar appear.}
\end{center}
\end{figure}

\begin{figure}[t]
\begin{center}
 \includegraphics[width=.48\textwidth]{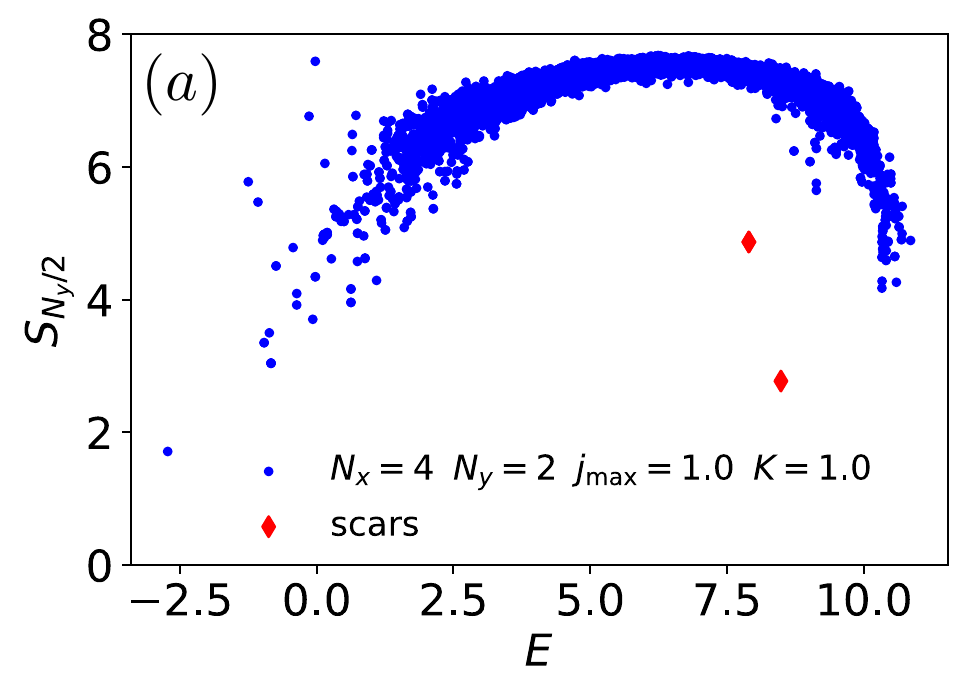}
 \includegraphics[width=.48\textwidth]{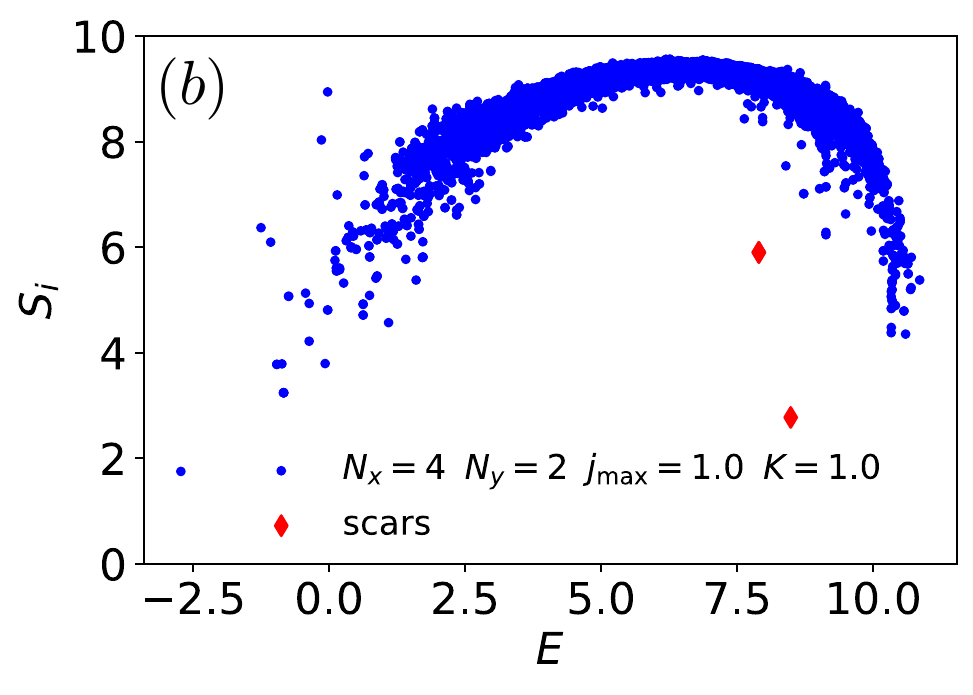}
\caption{\label{fig:EE_LWj1.0}(a) Bipartite entanglement entropy for all eigenvectors of the perturbed Levin-Wen at $j_{\rm max}=1.0$, $K=1.0$. (b) Shannon entropy for all eigenvectors of the perturbed Levin-Wen at $j_{\rm max}=1.0$, $K=1.0$.}
\end{center}
\end{figure}

In the previous subsection, we found that quantum scars appear in the Yang-Mills model~\eqref{eq:HYM}.
To gain a deeper understanding of these quantum scars, we investigate whether the Gauss-law constraints alone are sufficient for their existence.
For this purpose, we consider the perturbed Levin-Wen model~\eqref{eq:HLW_perturbed} that shares the same Gauss-law constraints as the Yang-Mills model.
We first compute the spectrum of the Levin-Wen model with $N_x=4$, $N_y=2$, and $j_\mathrm{max}=3/2$.
The dimension of the physical Hilbert space is summarized in table~\ref{table:dim}, where we restrict $j$ to integers ($j=0,1$). 
Under the constraint, the model is reduced to the Fibonacci anyons model (see e.g., ref.~\cite{Feiguin:2006ydp}).
Note that if $j_\mathrm{max} = 1/2$, the string-net model is identical to the Yang-Mills model, in which only the Wilson loop with the fundamental representation is allowed.
We show the bipartite entanglement entropy $S_{N_y/2}$ and the Shannon entropy $S_i$ in figure~\ref{fig:EEj1.5}. 
No quantum scars appear in the spectrum, and from this observation, we find that the dynamics is important as well as the constrained Hilbert space given by the Gauss law. We also found that no zero modes of the magnetic Hamiltonian that are simultaneously eigenvectors of the electric Hamiltonian appear in the perturbed Levin-Wen model.

Lastly, we compute the spectrum of the Levin-Wen model with $N_x=4$, $N_y=2$, and $j_\mathrm{max}=1$.
We show the bipartite entanglement entropy $S_{N_y/2}$ and the Shannon entropy $S_i$ in figure~\ref{fig:EE_LWj1.0}.
The number of scars is reduced from $29$ (Yang-Mills) to $3$ (Levin-Wen). We found that the scars are simultaneous eigenstates of the electric and magnetic Hamiltonians, but they have nonzero eigenvalue for the magnetic Hamiltonian (i.e., they are no longer zero modes). Although their energy changes as $K$ varies, but their wave functions and the expectation values are independent of $K$.

\section{Conclusions}
\label{sec:conclusions}

We have studied the Hamiltonian lattice Yang-Mills theory based on spin networks.
Using spin networks, we obtain the efficient graphical representations of physical states and action of the Kogut-Susskkind Hamiltonian to the physical states as summarized in eqs.~\eqref{eq:Hamiltonian},~\eqref{eq:action_electric_field} and~\eqref{eq:action_trU}, or eqs.~\eqref{eq:HYM},~\eqref{eq:action_electric_field_q} and~\eqref{eq:action_trU_q}.
To do numerical simulations, we regularized the theory based on the $q$ deformation.
This regularization respects the (discretized) gauge symmetry as quantum group, which enables implementation in both classical and quantum algorithms by referring to those of the string-net model. 
For example, a circuit implementation of the Wilson loop can be done in the same way as that in the string-net model (see e.g., refs.~\cite{PhysRevB.86.165113,PRXQuantum.3.040315,PhysRevX.12.021012}).
Such implementation will be elaborated in a future study.
Furthermore, we have studied quantum scars in a nonabelian gauge theory.
We found that quantum scars from zero modes arise even in a nonabelian gauge theory.
Comparison of the Yang-Mills model with the perturbed string-net model revealed that the presence of quantum scars is not guaranteed only by the Gauss law constraints.
We need to elaborate algebraic structures hidden in the Yang-Mills or Levin-Wen model to find a general scaling law of the number of quantum scars for changing the system size or cutoff $k$.
Since quantum scars are nonthermal states that break ergodicity, it is interesting to study the effects of quantum scars in thermalization of a small Yang-Mills system~\cite{Hayata:2020xxm}.

\vspace{1em}

\textbf{Note added:} 
While finalizing our manuscript, we became aware of a related work recently posted on arXiv~\cite{Zache:2023dko}. Their study proposes a similar method to regularize the infinite dimension of the Hilbert space of nonabelian gauge theories, and we acknowledge the concurrent efforts in this research field.


\section*{Acknowledgements}
The numerical calculations were carried out on cluster computers at iTHEMS in RIKEN.
This work was supported by JSPS KAKENHI Grant Numbers~21H01007, and 21H01084.

\appendix
\section{Single Plaquette Model}
\label{sec:appendix}
\begin{figure}[t]
  \begin{center}
   \includegraphics[width=.17\textwidth]{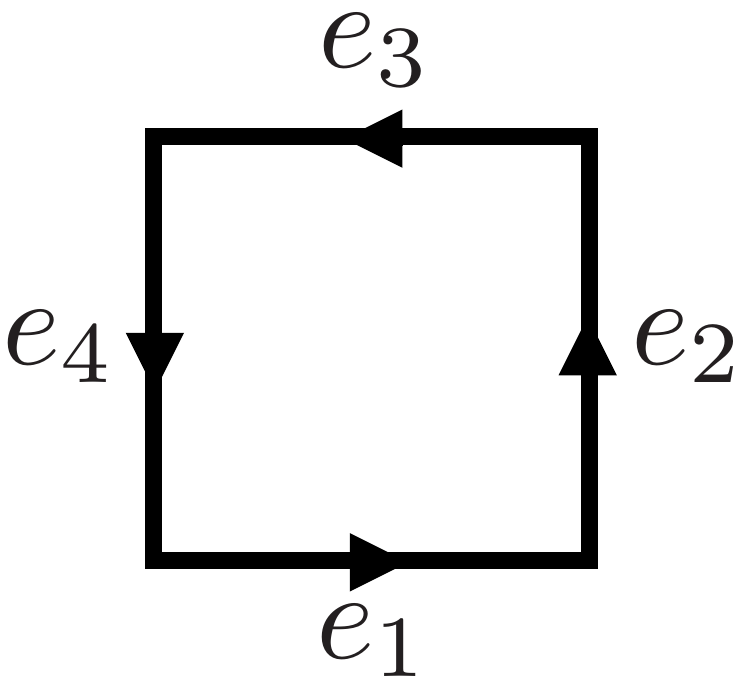}
  \caption{\label{fig:single_plaquette}Graph for a single plaquette model.}
  \end{center}
  \end{figure}
  
\begin{figure}[t]
\begin{center}
 \includegraphics[width=.7\textwidth]{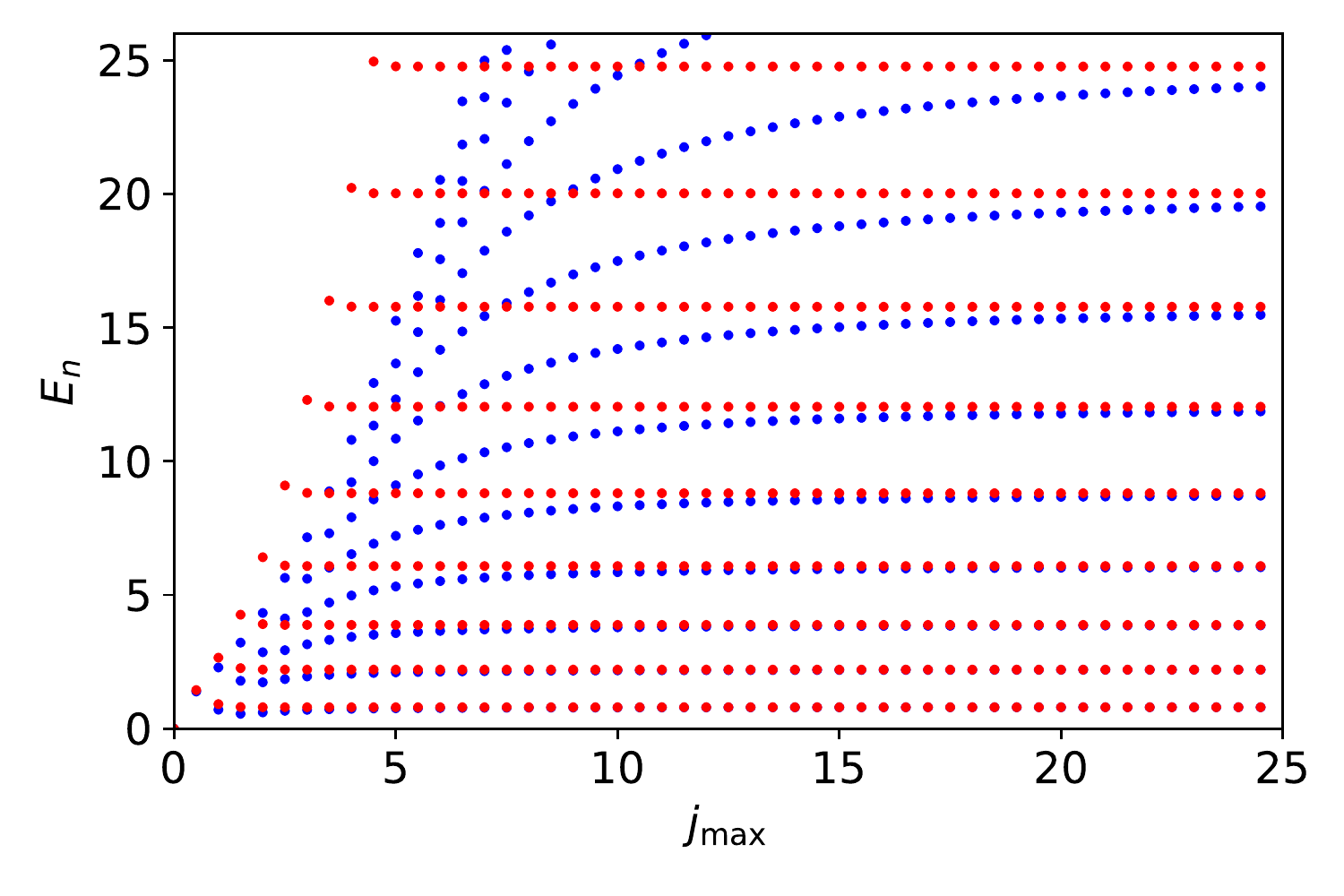}
\caption{\label{fig:Energy_su2}Energy eigenvalues $E_n$ of the single plaquette model for $\mathrm{SU}(2)_k$ [$j_{\rm max}=k/2$]. Red and blue dots represent the results of exact diagonalization with naive cutoff and $q$-deformation, respectively.}
\end{center}
\end{figure}
To see cutoff dependence and compare it with naive cutoff regularization, we compute the energy spectrum of a single plaquette model whose graph is shown in figure~\ref{fig:single_plaquette}.
We first consider $\mathrm{SU}(2)$ with and without the $q$-deformation.
The Gauss-law constraints require that all $j_{e_i}$'s are equal, so the basis can be represented by a single $j$ as $\ket{j}\coloneqq\ket{j_{e_1}j_{e_2}j_{e_3}j_{e_4}}$ with $j=j_{e_1}=j_{e_2}=j_{e_3}=j_{e_4}$.
The Hamiltonians in this basis is
\begin{equation}
  H = \sum^{j_\mathrm{max}}_{j=0} 2C_2(j) \ket{j}\bra{j}- K\sum^{j_\mathrm{max}-\frac{1}{2}}_{j=0} |j+\frac{1}{2}\rangle\langle j|- K\sum^{j_\mathrm{max}}_{j=1} |j-\frac{1}{2}\rangle\langle j| .
  \label{eq:Hamiltonian_one_plaquette_su2}
\end{equation}
Here, the factor of $2$($=(1/2)\times 4$) in front of $C_2(q)$ reflects there are four edges.
The $F$-symbol appeared in the Hamiltonian is
\begin{equation}
  [F_{j'}^{0j\frac{1}{2}}]_{jj'}=\delta_{j',j+1/2}+\delta_{j,j'+1/2},
\end{equation}
which leads to the matrix element of the magnetic term in eq.~\eqref{eq:Hamiltonian_one_plaquette_su2}.
In this model, the difference between the $q$-deformed and the naive cutoff regularization appears only in the quadratic Casimir invariant: $C_2(j)=[j][j+1]$ for the $q$-deformation and $C_2(j)=j(j+1)$ for the naive cutoff.

\begin{figure}[t]
  \begin{center}
   \includegraphics[width=.6\textwidth]{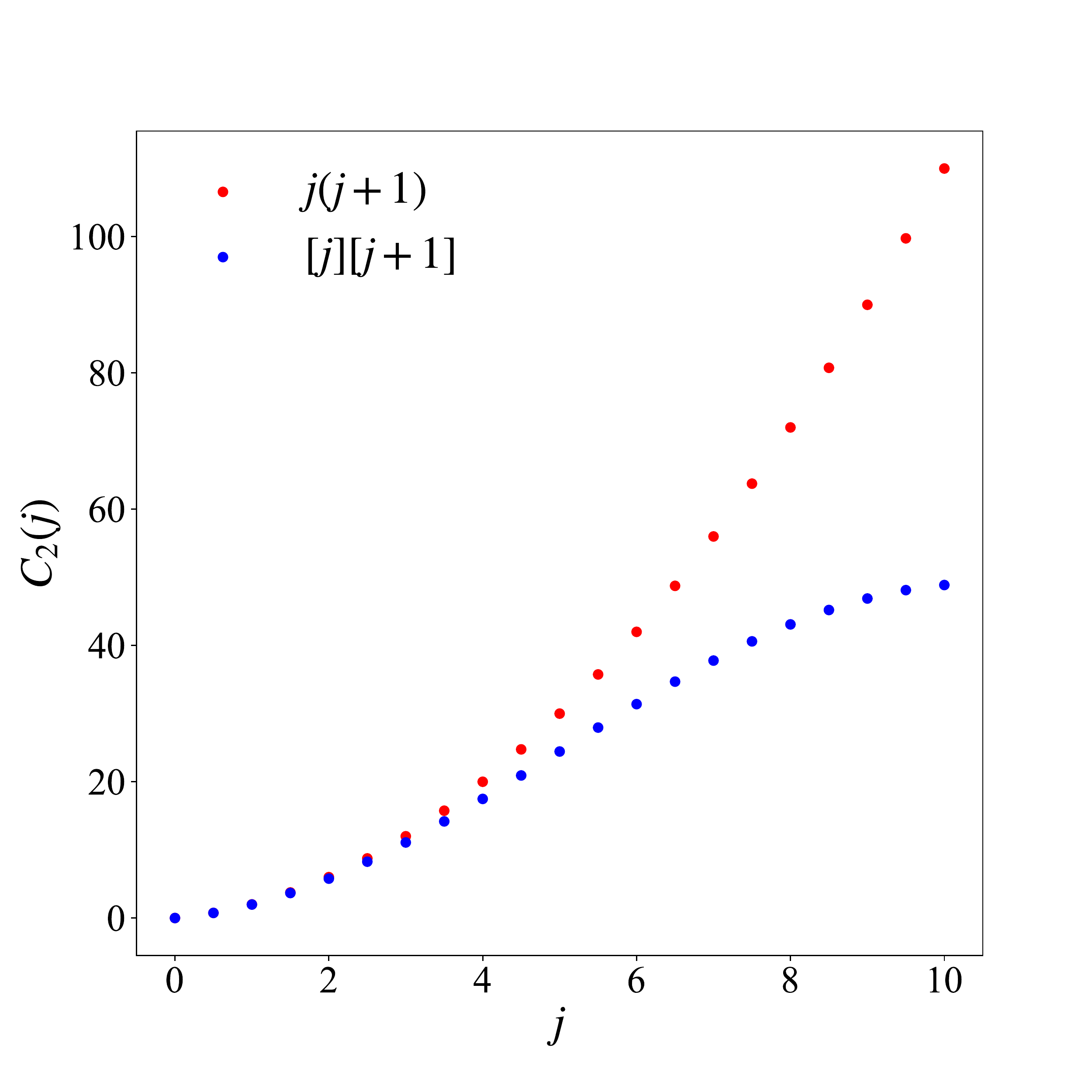}
  \caption{\label{fig:casimir}Quadratic Casimir invariant $C_2(j)$ for naive cutoff (red dots) and $q$-deformation (blue dots) for $j_\mathrm{max}=10$.}
  \end{center}
  \end{figure}

We show eigenvalues of the Hamiltonian with $K=1$ in figure~\ref{fig:Energy_su2}.
We see that lower eigenvalues become independent of $j_\mathrm{max}$ as $j_\mathrm{max}$ increases both in the naive cutoff and the $q$-deformation.
Thus the physics does not depend on the choice of cutoff as long as it is large enough.
However, we may need a ``better'' cutoff in practical computations to reduce computational costs.
We see that the naive cutoff has good convergence for energy eigenvalues compared with that based on the $q$-deformation as shown in figure~\ref{fig:Energy_su2}.
This is natural because the $q$-deformation softens the potential barrier by $C_2(j)$ as shown in figure~\ref{fig:casimir}, so that higher $j$ states are more excited, which may require larger $j_\mathrm{max}$.
However, this does not mean that the naive cutoff is better than the $q$-deformation.
While the $q$-deformation respects remnants of the continuous gauge symmetry as a quantum group after the discretization of the $\mathrm{SU}(2)$ manifold, the naive cutoff explicitly breaks the symmetry. There may be a trade-off between symmetry (i.e., handleability) and accuracy.

\begin{figure}[t]
\begin{center}
 \includegraphics[width=.4\textwidth]{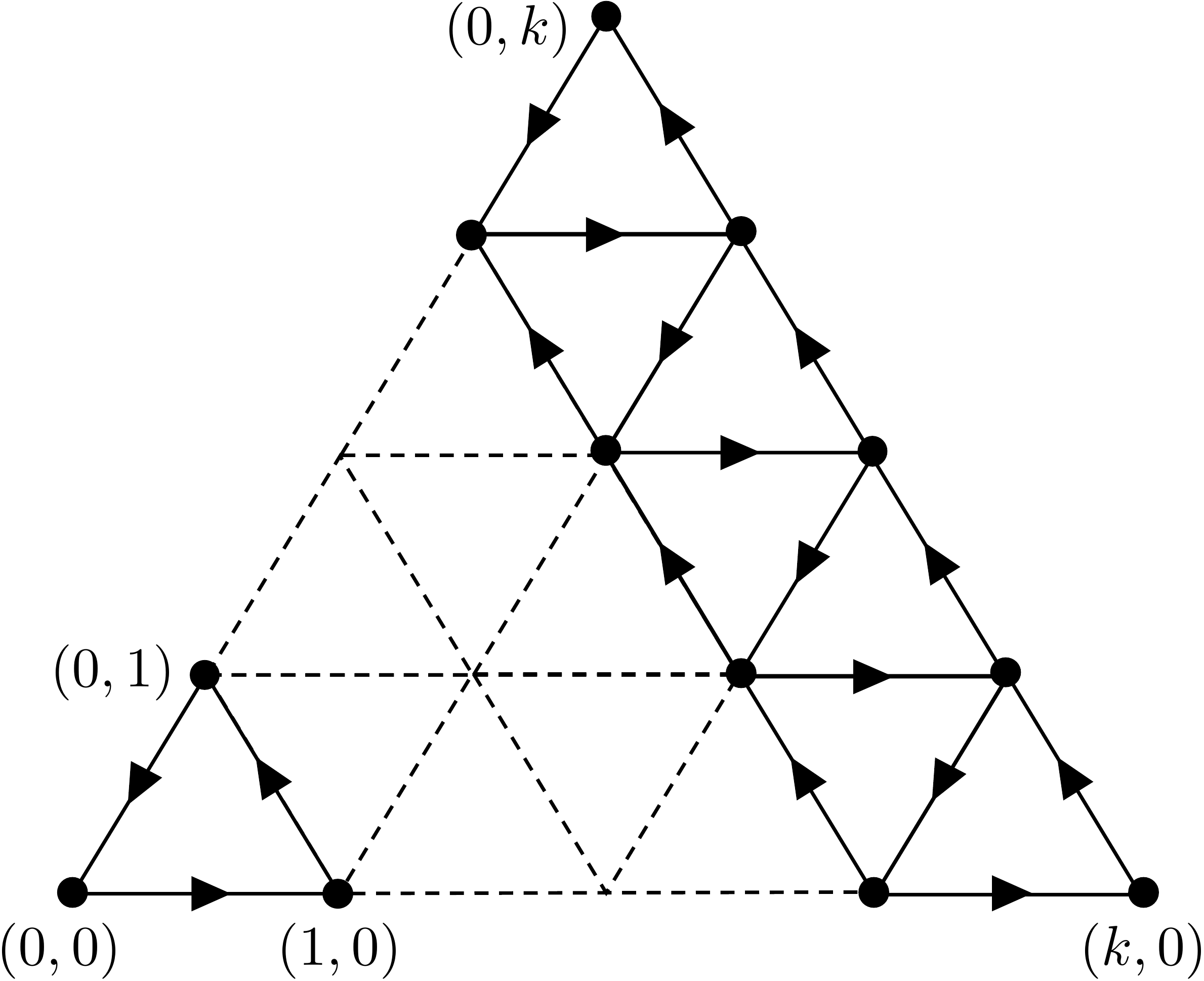}
\caption{\label{fig:boundary}Diagram for $\mathrm{SU}(3)_k$. Arrows indicate the action of the fundamental Wilson loop.}
\end{center}
\begin{center}
 \includegraphics[width=.7\textwidth]{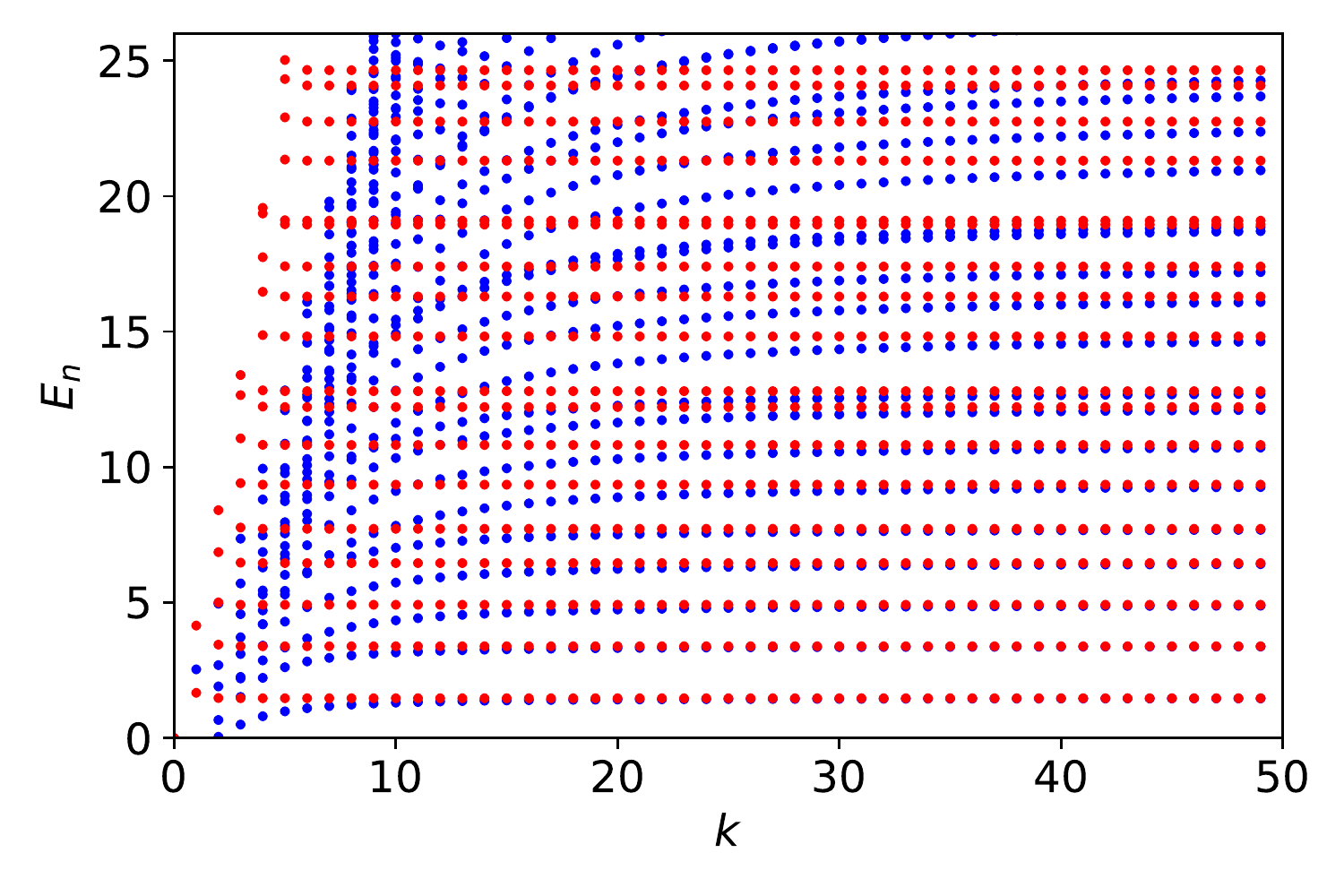}
\caption{\label{fig:Energy_su3}Energy eigenvalues $E_n$ of the single plaquette model for $\mathrm{SU}(3)_k$. Red and blue dots represent the results of exact diagonalization with naive cutoff and $q$-deformation, respectively.}
\end{center}
\end{figure}

Similarly, we can solve the single plaquette model for SU($3$) with $q$-deformation.
After solving the Gauss-law constraints, the basis of the single plaquette model is represented by pairs $(p,q)$ (the Dynkin index) as $|p,q\rangle$, where $p$ and $q$ are restricted to $0\leq p,q\leq k$ and $p+q\leq k$.
The Hamiltonian of $\mathrm{SU}(3)$ single plaquette model is
\begin{equation}
\begin{split}
  H &= \sum^{k}_{p=0}\sum^{k-p}_{q=0} 2C_2(p,q)\ket{p,q}\bra{p,q}
  \\
  &- K\sum^{k}_{p=0}\sum^{k-p}_{q=0}\qty( |p+1,q\rangle\langle p,q|+|p,q-1\rangle\langle p,q|+|p-1,q+1\rangle\langle p,q|)
\\
  &- K\sum^{k}_{p=0}\sum^{k-p}_{q=0}\qty( |p,q\rangle\langle p+1,q|+|p,q\rangle\langle p,q-1|+|p,q\rangle\langle p-1,q+1|),
  \label{eq:Hamiltonian_one_plaquette_su3}
\end{split}
\end{equation}
where the second order Casimir invariant $C_2(p,q)$ for $\mathrm{SU}(3)_k$ is given as (see, e.g., ref.~\cite{Bonatsos:1999xj}\footnote{We use the normalization factor of $C_2(p,q)$ commonly employed in high-energy physics, which is half of the value used in ref.~\cite{Bonatsos:1999xj}.}.)
\begin{equation}
  C_2(p,q)  = \frac{1}{2}\qty(\left[\frac{p}{3}-\frac{q}{3}\right]^2+\left[\frac{2p}{3}+\frac{q}{3}+1\right]^2+\left[\frac{p}{3}+\frac{2q}{3}+1\right]^2-2).
\end{equation}
See also, e.g., ref.~\cite{Ciavarella:2021nmj} for the action of the Wilson loop.

The cutoff to $(p,q)$ and boundary conditions to hopping terms induced by the Wilson loop operator are shown in figure~\ref{fig:boundary}.
We show eigenvalues of the Hamiltonian with $K=1$ in figure~\ref{fig:Energy_su3}.
As in the case of $\mathrm{SU}(2)$, the $q$-number $[p]$ is replaced as $[p]=p$ in the naive cutoff.
We see that lower eigenvalues become independent of cutoff $k$ as $k$ increases.
For energy eigenvalues, we see that naive cutoff has good convergence compared with that based on the $q$-deformation as shown in figure~\ref{fig:Energy_su3}.
Thus, qualitative behavior is the same as $\mathrm{SU}(2)_k$.

\bibliographystyle{JHEP}
\bibliography{bib}

\end{document}